\documentclass{aa}  

\usepackage{graphicx}
\usepackage{txfonts}

\usepackage{bm}
\usepackage{comment}
\usepackage{mathdots}
\usepackage{multirow}
\usepackage{textcomp}
\usepackage{ulem}
\usepackage{url}
\usepackage{xcolor}
\begin{document}

   \title{Increasing the achievable contrast of infrared interferometry with an error correlation model\thanks{Based on observations made with ESO telescopes at Paranal Observatory under programme IDs 60.A-9801(U) and 0101.C-0907(B).}}


   \author{J. Kammerer
          \inst{1,2}
          \and
          A. M\'erand\inst{1}
          \and
          M. J. Ireland\inst{2}
          \and
          S. Lacour\inst{1,3}
          }
   
   \institute{European Southern Observatory, Karl-Schwarzschild-Str. 2, 85748, Garching, Germany \\
              \email{jens.kammerer@eso.org}
         \and
             Research School of Astronomy \& Astrophysics, Australian National University, ACT 2611, Australia
         \and
LESIA, Observatoire de Paris, Université PSL, CNRS, Sorbonne Université,
Université de Paris, 5 place Jules Janssen, 92195 Meudon, France
             }

   \date{Received September 15, 1996; accepted March 16, 1997}

 
  \abstract
  {Interferometric observables are strongly correlated, yet it is common practice to ignore these correlations in the data analysis process.}
   {We develop an empirical model for the correlations present in Very Large Telescope Interferometer GRAVITY data and show that properly accounting for them yields fainter detection limits and increases the reliability of potential detections.}
   {We extracted the correlations of the (squared) visibility amplitudes and the closure phases directly from intermediate products of the GRAVITY data reduction pipeline and fitted our empirical models to them. Then, we performed model fitting and companion injection and recovery tests with both simulated and real GRAVITY data, which are affected by correlated noise, and compared the results when ignoring the correlations and when properly accounting for them with our empirical models.}
   {When accounting for the correlations, the faint source detection limits improve by a factor of up to $\sim 2$ at angular separations $> 20~\text{mas}$. For commonly used detection criteria based on $\chi^2$ statistics, this mostly results in claimed detections being more reliable.}
   {Ignoring the correlations present in interferometric data is a dangerous assumption which might lead to a large number of false detections. The commonly used detection criteria (e.g. in the model fitting pipeline CANDID) are only reliable when properly accounting for the correlations; furthermore, instrument teams should work on providing full covariance matrices instead of statistically independent error bars as part of the official data reduction pipelines.}


   \maketitle
%

\section{Introduction}
\label{sec:introduction}

With the first detection and characterisation of an exoplanet by the Very Large Telescope Interferometer (VLTI) instrument GRAVITY \citep[HR 8799 e,][]{gravity2019}, infrared interferometry has proven to be a powerful technique for high-contrast imaging at high angular resolution. Although initially designed for observations of the galactic centre \citep{bartko2009}, GRAVITY's dual-feed mode combined with the recently installed integrated optics beam combiner \citep{perraut2018} enable spectroscopy and micro-arcsecond astrometry of exoplanets with a wide range of angular separations \citep{gravity2019}.

More recently, \citet{gravity2020} have used GRAVITY observations of $\beta$ Pic b in order to derive reliable estimates for the mass and the C/O ratio of the young giant planet using forward modelling and free retrieval of its atmosphere. In the future, infrared interferometry will be a promising opportunity for studying giant planet formation \citep[e.g. with Hi-5,][]{defrere2018} and potentially even characterising terrestrial exoplanets from space \citep[e.g. with a formation-flying nulling interferometer,][]{leger1996,mennesson1997,kammerer2018, quanz2018, quanz2019}. However, significant improvements are required on the technical side \citep[e.g. kernel nulling,][]{martinache2018}, in addition to on the data reduction side in order to achieve these ambitious goals.

Because they use the dual-feed mode of GRAVITY, the aforementioned observations are not conducted anywhere close to the diffraction limit of the interferometer, but rather the diffraction limit of a single telescope. Detecting a companion within the interferometer’s diffraction limit (a few $\lambda/b_\text{max}$), where $\lambda$ is the observing wavelength and $b_\text{max}$ is the longest baseline of the interferometer, is limited by systematic errors. While such systematic errors that are introduced by instrumental and atmospheric effects have been studied intensively (e.g. imperfect fibre coupling, \citealt{kotani2003}; instrument vibrations, \citealt{lebouqin2011}; differential atmospheric piston; \citealt{colavita1999}), correlations are also introduced by the data reduction and the calibration. For instance, a systematic error might be introduced similarly to all complex visibilities measured on the science target if the instrumental transfer function obtained from the calibrator target is affected by unknown variability \citep{perrin2003} and if the closure phases measured over telescope triplets of closing triangles are not mathematically independent \citep{monnier2007}. Nevertheless, most data reduction pipelines (e.g. the PIONIER data reduction pipeline, \citealt{lebouqin2011}; the GRAVITY data reduction pipeline, \citealt{lapeyrere2014}) and model fitting routines (e.g. LITpro\footnote{\url{https://www.jmmc.fr/english/tools/data-analysis/litpro}}, \citealt{tallon-bosc2008}; CANDID\footnote{\url{https://ascl.net/1505.030}}, \citealt{gallenne2015}) assume statistically independent observables. However, in order to robustly detect faint companions, or place upper limits on their brightness, a solid understanding and description of the systematic errors is inevitable.

While \citet{lachaume2019} proposed to use the bootstrapping method \citep[i.e. sampling with replacement,][]{efron1986} in order to obtain the multivariate probability density function of the squared visibility amplitudes and the closure phases, \citet{gravity2020} extracted the covariances of the complex visibilities directly from the data. Although the bootstrapping method is computationally expensive, it enables estimating the systematic errors not only between the different spectral channels, baselines and triangles, but also between different observations. This enables accounting for correlations introduced by sky rotation or the calibration method, but is only applicable at a higher level when the structure of the observing sequence is known.

In this paper, we follow a similar approach to \citet{gravity2020} by extracting the correlations between the squared visibility amplitudes and the closure phases directly from the data. Then, we develop an empirical model for these correlations which can be fitted to the correlations extracted from single GRAVITY pipeline products, even if only a small number of measurements is available. This enables the attainment of a systematic error estimate for every GRAVITY data set and could ultimately be included in the GRAVITY data reduction pipeline \citep{lapeyrere2014}.


\section{Methods}
\label{sec:methods}

In Section~\ref{sec:correlations_extracted_from_gravity_data} we show how we extract the correlations from individual GRAVITY pipeline products and describe their nature. In Section~\ref{sec:empirical_model_for_the_correlations} we introduce our empirical model for these correlations and in Section~\ref{sec:model_fitting} we present the model fitting routines with the aid of which we show the improvements that come from using our empirical correlation model. The Python code that we developed in the scope of this paper is publicly available on GitHub (\url{https://github.com/kammerje/InterCorr}).

\subsection{Correlations extracted from GRAVITY data}
\label{sec:correlations_extracted_from_gravity_data}

In order to extract the correlations between the different spectral channels, baselines and triangles from GRAVITY data we use the P2VM-reduced files from the GRAVITY data reduction pipeline. These files are intermediate pipeline products which contain the individual measurements (detector read-outs) before they are averaged together. Having access to the individual measurements enables extracting the correlations from the (complex) coherent flux $\text{VIS}_{mb\lambda}$ which is stored in the P2VM-reduced file as a data cube of shape $M \times B \times \Lambda$, where $m = 1..M$ is the number of individual measurements, $b = 1..B$ is the number of baselines and $\lambda = 1..\Lambda$ is the number of spectral channels. From the coherent flux, we compute the squared visibility amplitudes
\begin{equation}
    \text{VIS2}_{mb\lambda} = \frac{|\text{VIS}_{mb\lambda}|^2}{\text{F1F2}_{mb\lambda}},
\end{equation}
where $\text{F1F2}_{mb\lambda}$ is the product of the total fluxes, and the closure phases
\begin{equation}
    \text{T3}_{mt\lambda} = \bm{K} \cdot \angle\text{VIS}_{mb\lambda},
\end{equation}
where $t = 1..T$ is the number of triangles, $\bm{K}$ is a stack of $M$ matrices $\bm{k}$ which encode how the four unique triangles can be formed from the six unique baselines of the VLTI, that is
\begin{equation}
    \bm{k} = \begin{pmatrix}
    1 & -1 & 0 & 1 & 0 & 0 \\
    1 & 0 & -1 & 0 & 1 & 0 \\
    0 & 1 & -1 & 0 & 0 & 1 \\
    0 & 0 & 0 & 1 & -1 & 1
    \end{pmatrix},
\end{equation}
and $\angle$ denotes the argument of a complex number (i.e. the phase).

Then, we compute the sample covariance of the squared visibility amplitudes and the closure phases according to
\begin{equation}
    \label{eqn:covariance_matrix}
    (\Sigma_X)_{ij} = \frac{1}{M-1}\sum^M_{m=1}(X_{mi}-\bar{X}_i)(X_{mj}-\bar{X}_j),
\end{equation}
where $X$ is VIS2/T3 and $i$ and $j$ run over $1..B\Lambda$/$1..T\Lambda$ so that we obtain covariance matrices of shape $(B\Lambda) \times (B\Lambda)$/$(T\Lambda) \times (T\Lambda)$ that contain the covariances between the different spectral channels and baselines/triangles. $\bar{X}$ denotes the mean of $X$ over the individual measurements, that is
\begin{equation}
    \bar{X}_i = \frac{1}{M}\sum^M_{m=1}X_{mi}.
\end{equation}
The correlations between the VIS2 and the T3 then follow by dividing the covariances by the standard deviation $\sigma_i = \sqrt{\Sigma_{ii}}$ of the corresponding observables, that is
\begin{equation}
    \label{eqn:correlation_matrix}
    (C_X)_{ij} = \frac{(\Sigma_X)_{ij}}{(\sigma_X)_i(\sigma_X)_j}.
\end{equation}
The diagonal of the covariance matrix $\bm{\Sigma}$ equals the square of the standard deviation and that the diagonal of the correlation matrix $\bm{C}$ equals one by definition.

For developing an empirical correlation model we use data taken with GRAVITY \citep{gravity2017} at the Very Large Telescope Interferometer (VLTI) during technical time (programme 60.A-9801(U)). GRAVITY operates in the K-band ($2.0$--$2.4~\text{\textmu m}$) and combines the light from either the four Unit Telescopes (UTs) or the four Auxiliary Telescopes (ATs) of the VLTI in order to perform interferometric imaging and astrometry by phase referencing\footnote{\url{https://www.eso.org/sci/facilities/paranal/instruments/gravity.html}}. The data used here was taken with the four UTs on the object HD 82383 ($\zeta$ Ant B) in single-field medium resolution ($R = \lambda/\Delta\lambda \approx 500$) mode. This object is relatively bright (K-band magnitude of 6.698, cf. SIMBAD\footnote{\url{http://simbad.u-strasbg.fr/simbad/}}), hence the short integration time of $0.85~\text{ms}$ for the fringe tracker and $1~\text{s}$ for the science camera. It has a companion at an angular separation of $\sim 8~\text{arcsec}$ (cf. WDS\footnote{\url{http://www.astro.gsu.edu/wds/}}) which is well beyond the interferometric field-of-view. By choosing a bright target with a short exposure time we make sure that there is a sufficient number of frames to compute the sample covariance (cf. Equation~\ref{eqn:covariance_matrix}). The short exposure time of the fringe tracker \citep[much less than the atmospheric coherence time $t_0$, which is typically $\sim 20~\text{ms}$ in the K-band,][]{kellerer2007} and spatially filtered nature of the GRAVITY beam combiner means that the fringe tracker data is less affected by systematic errors than many other beam combiners. We extract the correlations from a single P2VM-reduced file (GRAVI.2019-03-29T02-01-37.193\_singlecalp2vmred.fits) in order to demonstrate the direct applicability of our method to the GRAVITY data reduction pipeline. Correlations extracted from other P2VM-reduced files of the same program can be found in the Appendix (Figure~\ref{fig:z_corr_SC}).

\begin{figure*}
\begin{center}
\includegraphics[width=\textwidth]{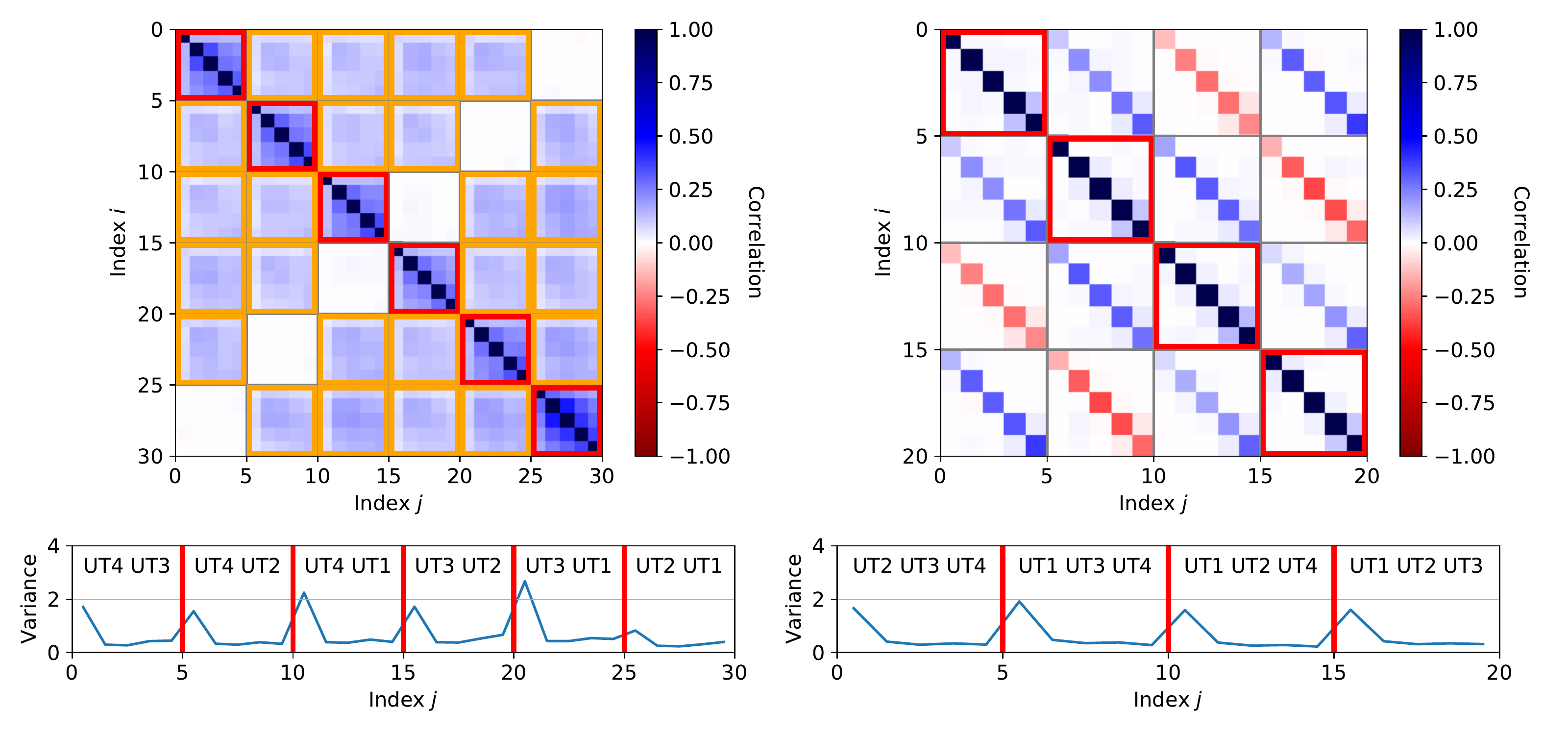}
\end{center}
\caption{Correlations of the VIS2 (left panel) and the T3 (right panel) for the GRAVITY fringe tracker, extracted from a single P2VM-reduced file. The axes run over the different baselines/triangles, with each individual baseline/triangle comprising five spectral channels. Correlations within the same baseline/triangle are highlighted with red squares and correlations between baselines having a telescope in common are highlighted with orange squares. We note that the correlations are computed from 46592 individual measurements. Below each panel, the variance of the data and the names of the telescopes forming each baseline/triangle are shown.}
\label{fig:corr_FT}
\end{figure*}

Figure~\ref{fig:corr_FT} shows the correlations of the VIS2 (left panel) and the T3 (right panel) for the GRAVITY fringe tracker. There are six different baselines and four different triangles with five spectral channels each, so 30 observables for the VIS2 and 20 observables for the T3 in total. Correlations within the same baseline/triangle are highlighted with red squares and correlations between baselines having a telescope in common are highlighted with orange squares.

The most dominant correlations of the VIS2 are between different spectral channels within the same baseline, with neighbouring spectral channels being affected most strongly. We suspect that these correlations are predominantly of both atmospheric or instrumental origin, since all five spectral channels follow the same optical path through the atmosphere and up to the dispersive element behind the beam combiner and before the science camera. Also, the five spectral channels do not correspond to individual pixels on the detector of the fringe tracker. In fact, the wavelengths of the five spectral channels lie somewhere between the wavelengths corresponding to the pixels on the detector of the fringe tracker, so that the values recorded by two neighbouring pixels on the detector need to be interpolated in order to find the values for the five spectral channels of the fringe tracker. This could explain the strong correlations between neighbouring spectral channels (one pixel above or below the diagonal) observed for the VIS2, but also for the T3. Furthermore, there are significant correlations between baselines having a telescope in common. Their strength is roughly half the strength of the correlations within the same baseline, which makes sense if the correlations are introduced by atmospheric or instrumental effects and affect each of the four individual beams of the interferometer separately. Also, baselines having no telescope in common are essentially uncorrelated. Hence, we conclude that most of the correlations of the VIS2 are caused by atmospheric or instrumental effects.

For the T3, we observe similar correlations between neighbouring spectral channels as for the VIS2. This makes sense since the closure phases are built from a linear combination (encoded in the matrix $\bm{k}$) of the phase of the complex visibilities, whose absolute square are the squared visibility amplitudes. Moreover, there are significant correlations of $\sim \pm 1/3$ between the same spectral channels on different triangles. These are caused by the fact that each set of two different triangles has exactly one of their three baselines in common, that is each column of the matrix $\bm{k}$ has exactly two non-zero entries. If the common baseline is shared between the different triangles in parallel direction (i.e. the two entries in the corresponding column of the matrix $\bm{k}$ have the same sign), the correlation is $+1/3$, otherwise it is $-1/3$. This structure with the side-diagonals being $\pm 1/3$ can also be explained by assuming uncorrelated visibility phases (i.e. a diagonal correlation matrix
\begin{equation}
    \bm{C}_{\angle\text{VIS}} = \begin{pmatrix}
    1 & 0 & \cdots & 0 \\
    0 & \ddots & \ddots & \vdots \\
    \vdots & \ddots & \ddots & 0 \\
    0 & \cdots & 0 & 1
    \end{pmatrix}
\end{equation}
of shape $B\times\Lambda$ for the visibility phases) and performing a basis transform
\begin{equation}
    \bm{C}_\text{T3} = \bm{T} \cdot \bm{C}_{\angle\text{VIS}} \cdot \bm{T}^T,
\end{equation}
where $\bm{T}$ represents a matrix of shape $(T\Lambda) \times (B\Lambda)$ which maps the vector of visibility phases to the vector of closure phases and can be trivially obtained from the matrix $\bm{k}$. Also, the observed correlations between neighbouring spectral channels on different triangles (pixels next to the side-diagonals) are naturally explained by this basis transform given the correlations of the VIS2 observed between neighbouring spectral channels on the same baseline.

\begin{figure*}
\begin{center}
\includegraphics[width=\textwidth]{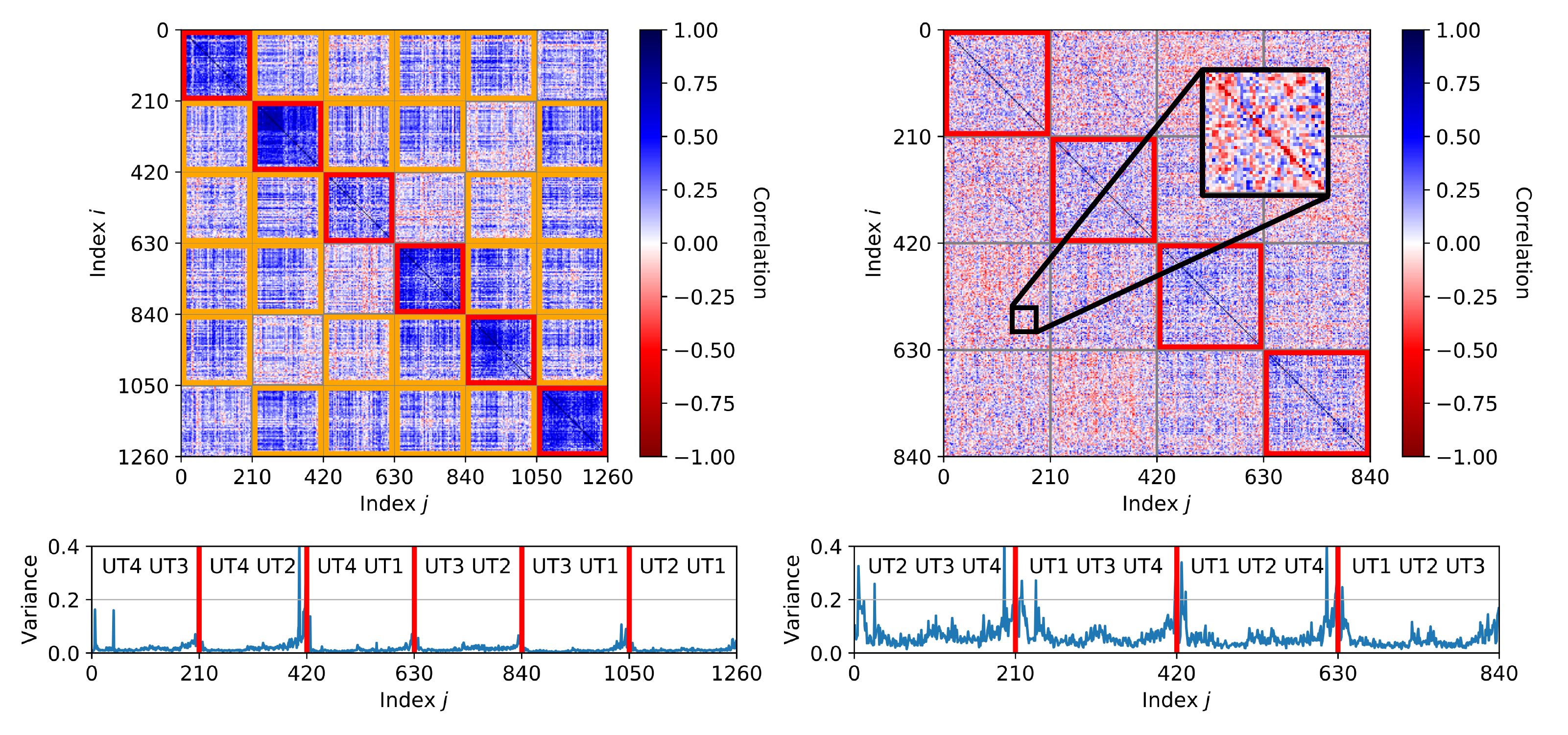}
\end{center}
\caption{Same as Figure~\ref{fig:corr_FT}, but for the GRAVITY science camera. Each individual baseline/triangle comprises 210 spectral channels. We note that the correlations are computed from 32 individual measurements.}
\label{fig:corr_SC}
\end{figure*}

Figure~\ref{fig:corr_SC} shows the correlations of the VIS2 (left panel) and the T3 (right panel) for the GRAVITY science camera. There are six different baselines and four different triangles with 210 spectral channels each, so 1260 observables for the VIS2 and 840 observables for the T3 in total.

Due to the much smaller number of individual measurements if compared to the fringe tracker the correlations of the science camera are more dominated by noise. Nevertheless, we observe strong positive correlations between different spectral channels within the same baseline (i.e. inside the red squares) and significant positive correlations between baselines having a telescope in common (i.e. inside the orange squares) for the VIS2, similar to the correlations observed for the fringe tracker. Although the atmospheric turbulence and the optical elements (i.e. mirrors, delay lines, optical fibres, beam combiner) seen by the science camera are similar to those seen by the fringe tracker, the exposure time of the science camera is much longer than both the atmospheric coherence time $t_0$ and the fringe tracker inverse $3~\text{dB}$ bandwidth \citep{lacour2019}, which means that the VIS2 correlations are expected to be decreased by a term proportional to the square of the fringe tracking error and the closure phase random errors are expected to be proportional to the cube of the fringe tracking error \citep{ireland2013}. Therefore, since there still are significant correlations for the science camera, they must be introduced by the (correlated) fringe tracker, forwarding the correlations shown in Figure~\ref{fig:corr_FT} to the science camera. For the T3, we again observe significant correlations of $\sim \pm 1/3$ between the same spectral channels on different triangles. On top of this, there are also weak positive correlations between different spectral channels on the same triangle (i.e. inside the red squares) and between different spectral channels on different triangles whose sign depends on whether the corresponding triangles share a baseline in parallel or anti-parallel direction. Again, these correlations are naturally explained by the basis transform $\bm{T}$ given the correlations observed for the VIS2 of the science camera.

\subsection{Empirical model for the correlations}
\label{sec:empirical_model_for_the_correlations}

An empirical VIS2/T3 sample covariance with fewer frames than the product of the number of baselines/triangles and spectral channels is necessarily singular. It takes a number of frames much greater than this to estimate a sample covariance matrix with a condition number approaching that of the true sample covariance. For this reason, we choose to develop an analytic model for the covariance matrix $\bm{\Sigma}$ of the VIS2 and the T3. This model can then be fitted to the (potentially under-conditioned) covariance extracted from an arbitrary GRAVITY data set and can be used for model fitting based on log-likelihood maximisation. Most model fitting routines (e.g. LITpro, \citealt{tallon-bosc2008}; CANDID, \citealt{gallenne2015}) are based on $\chi^2$ minimisation, which is equivalent to log-likelihood maximisation, where
\begin{equation}
    \chi^2 = R^T \cdot \bm{\Sigma}^{-1} \cdot R
\end{equation}
and $R = D-M$ is the residual between data and model (cf. Section~\ref{sec:model_fitting}).

Our approach is to model the correlation matrices $\bm{C}_\text{VIS2}$ of the VIS2 and $\bm{C}_\text{T3}$ of the T3 which have the relatively simple structure observed in Figures~\ref{fig:corr_FT} and~\ref{fig:corr_SC}. Moreover, the observed structure of the correlations is consistent for different data sets with different exposure times (1 s with the UTs for programme 60.A-9801(U) and 10 s with the ATs for programme 0101.C-0907(B), cf. also Figures~\ref{fig:z_corr_SC} and~\ref{fig:y_corr_SC}). Then, we compute
\begin{equation}
    \label{eqn:corr2cov}
    \Sigma_{ij} = C_{ij}\sigma_i\sigma_j,
\end{equation}
where $\sigma$ denotes the standard deviation of the data which can be obtained from the VIS2ERR and the T3PHIERR columns of the OIFITS files for example. We note that these standard deviations are used to build diagonal covariance matrices in LITpro and CANDID which assume uncorrelated data only. Of course, assuming uncorrelated data is a simplification and we discuss the problems that arise from this in Section~\ref{sec:injection_and_recovery_tests_simulated_data}.

A very important point is that Equation~\ref{eqn:corr2cov} only holds if the errors on the VIS2 ($\sigma_\text{VIS2}$) and the T3 ($\sigma_\text{T3}$) are reliably estimated by the GRAVITY data reduction pipeline. The pipeline manual\footnote{\url{http://www.eso.org/sci/software/pipelines/index.html#pipelines_table}} explains that the uncertainties are computed by bootstrapping over $\sim 10$ independent samples, so that the final error on the mean measurement is estimated from the observed statistics at a slightly higher temporal frequency. There is no re-scaling or accounting for systematics in this process. In case there are less than five frames available, Monte-Carlo realisations of the theoretical photon and detector noise are added to the samples, which leads to less realistic uncertainties. However, our data sets consist of 32 frames exposures for programme 60.A-9801(U) and 20 frames exposures for programme 0101.C-0907(B), respectively. While we understand that the use of the pipeline uncertainties is a limitation and that an incorrect noise model can reduce the detection sensitivity or yield false positives \citep[cf. e.g. Section~3 of][]{delisle2020}, we also note that investigating and quantifying the credibility of these uncertainties is beyond the scope of this work.

Our models for the correlation matrices equal one on the diagonal according to the definition of a correlation matrix (cf. Equation~\ref{eqn:correlation_matrix}, that is every observable is 100\% correlated with itself) and have one free parameter which can be determined by fitting the model to the correlations extracted from the P2VM-reduced files. For the correlation matrix of the VIS2 $\bm{C}_\text{VIS2}$, the free parameter $x$ represents the correlations between spectral channels within the same baseline and between baselines having a telescope in common. There are correlations of $x$ between different spectral channels within the same baseline, correlations of $x/2$ between baselines having a telescope in common, and no correlations between baselines having no telescope in common (cf. left panel of Figure~\ref{fig:corr_SC}), that is
\begin{align}
    \bm{C}_\text{VIS2} &= \begin{pmatrix}
    \bm{X}_1 & \bm{X}_2 & \cdots & \cdots & \cdots & \bm{X}_2 & 0 \\
    \bm{X}_2 & \ddots & \ddots & & \iddots & \iddots & \bm{X}_2 \\
    \vdots & \ddots & \ddots & \bm{X}_2 & 0 & \iddots & \vdots \\
    \vdots & & \bm{X}_2 & \ddots & \bm{X}_2 & & \vdots \\
    \vdots & \iddots & 0 & \bm{X}_2 & \ddots & \ddots & \vdots \\
    \bm{X}_2 & \iddots & \iddots & & \ddots & \ddots & \bm{X}_2 \\
    0 & \bm{X}_2 & \cdots & \cdots & \cdots & \bm{X}_2 & \bm{X}_1
    \end{pmatrix}, \\
    \bm{X}_1 &= \begin{pmatrix}
    1 & x & \cdots & x \\
    x & \ddots & \ddots & \vdots \\
    \vdots & \ddots & \ddots & x \\
    x & \cdots & x & 1
    \end{pmatrix}, \\
    \bm{X}_2 &= \begin{pmatrix}
    x/2 & \cdots & x/2 \\
    \vdots & \ddots & \vdots \\
    x/2 & \cdots & x/2
    \end{pmatrix}.
\end{align}
The correlation matrix is a block matrix consisting of $B \times B$ blocks, where each individual block is a $\Lambda \times \Lambda$ matrix. For the correlation matrix of the T3 $\bm{C}_\text{T3}$, the free parameter $y$ represents the correlations between spectral channels within the same triangle. Moreover, as illustrated by the basis transform $\bm{T}$, this naturally leads to correlations of $\pm 1/3$ between the same spectral channel of different triangles and $\pm y/3$ between different spectral channels of different triangles (cf. right panel of Figure~\ref{fig:corr_SC}), that is
\begin{align}
    \bm{C}_\text{T3} &= \begin{pmatrix}
    \bm{Y}_1 & \bm{Y}_2 & \cdots & \bm{Y}_2 \\
    \bm{Y}_2 & \ddots & \ddots & \vdots \\
    \vdots & \ddots & \ddots & \bm{Y}_2 \\
    \bm{Y}_2 & \cdots & \bm{Y}_2 & \bm{Y}_1
    \end{pmatrix}, \\
    \bm{Y}_1 &= \begin{pmatrix}
    1 & y & \cdots & y \\
    y & \ddots & \ddots & \vdots \\
    \vdots & \ddots & \ddots & y \\
    y & \cdots & y & 1
    \end{pmatrix}, \\
    \bm{Y}_2 &= \begin{pmatrix}
    \pm 1/3 & \pm y/3 & \cdots & \pm y/3 \\
    \pm y/3 & \ddots & \ddots & \vdots \\
    \vdots & \ddots & \ddots & \pm y/3 \\
    \pm y/3 & \cdots & \pm y/3 & \pm 1/3
    \end{pmatrix}.
\end{align}
The correlation matrix is a block matrix consisting of $T \times T$ blocks, where each individual block is a $\Lambda \times \Lambda$ matrix. The sign is positive if the two triangles share a baseline in parallel direction and negative if they share a baseline in anti-parallel direction.

\begin{figure*}
\begin{center}
\includegraphics[width=\textwidth]{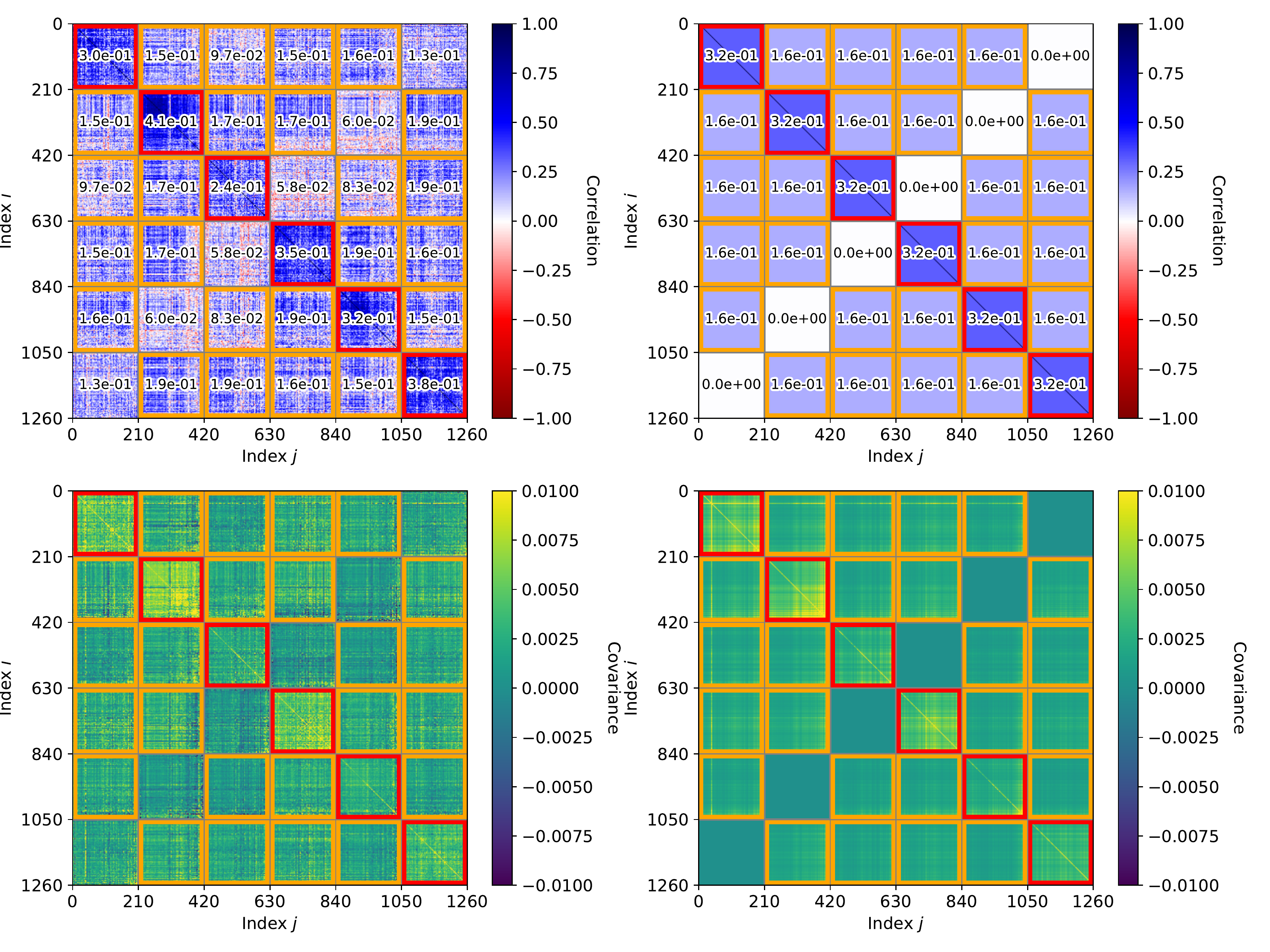}
\end{center}
\caption{Correlations of the VIS2 for the GRAVITY science camera, extracted from a single P2VM-reduced file (upper left panel) and our one-parameter model fitted to them (upper right panel). The bottom panels show the corresponding covariances obtained by multiplying the correlation $C_{ij}$ with the product of the standard deviations $\sigma_i\sigma_j$. Correlations/covariances within the same baseline are highlighted with red squares and correlations/covariances between baselines having a telescope in common are highlighted with orange squares.}
\label{fig:vis2fit_SC}
\end{figure*}

\begin{figure*}
\begin{center}
\includegraphics[width=\textwidth]{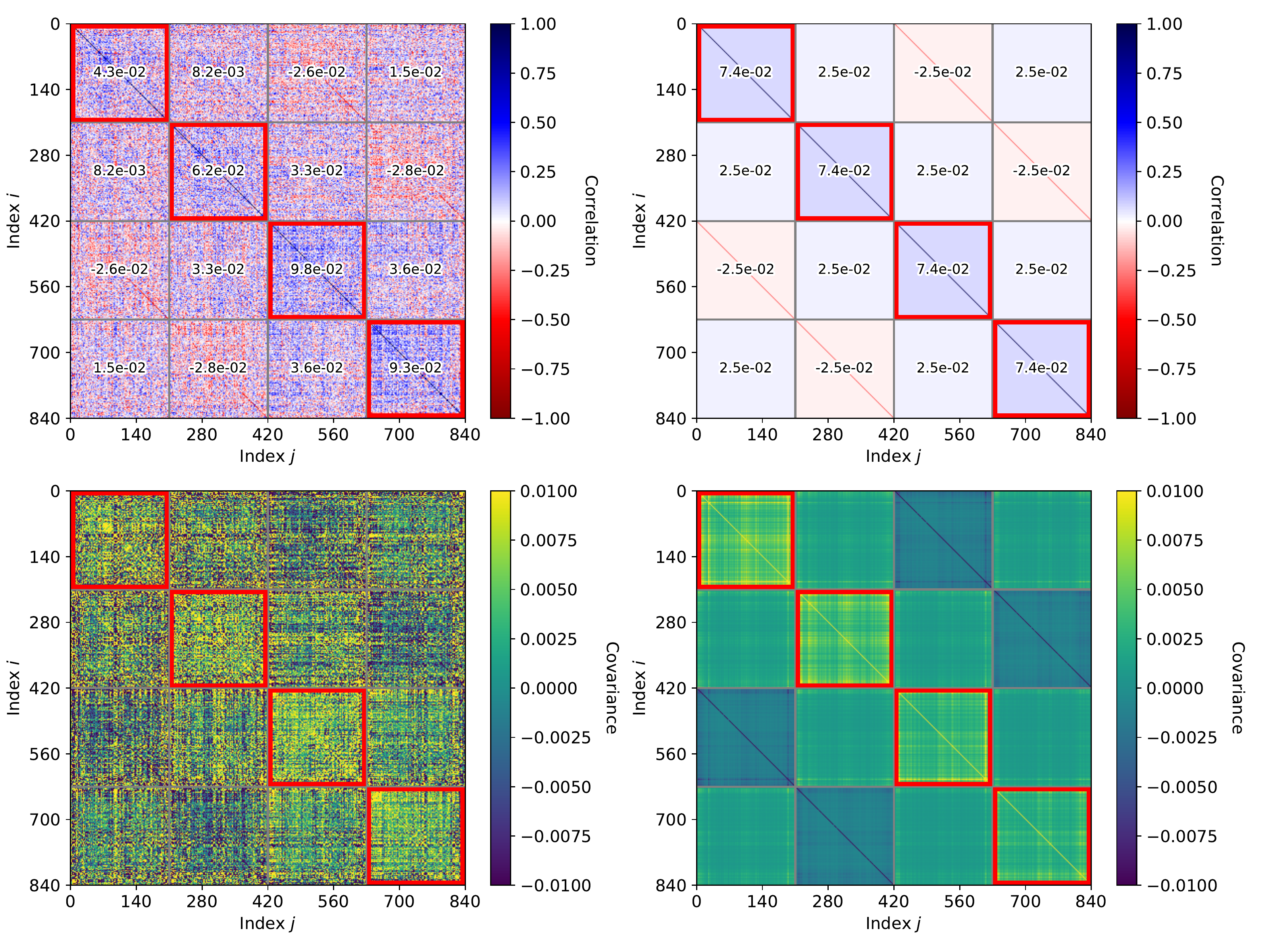}
\end{center}
\caption{Same as Figure~\ref{fig:vis2fit_SC}, but showing the correlations/covariances of the T3 and our one-parameter model fitted to them for the GRAVITY science camera.}
\label{fig:t3fit_SC}
\end{figure*}

We fit the previously described model to the correlations of the VIS2 and the T3 which we extracted from the single P2VM-redcued file of GRAVITY introduced in Section~\ref{sec:correlations_extracted_from_gravity_data}. Figure~\ref{fig:vis2fit_SC} shows the extracted and the model correlations (top panels) and the extracted and the model covariances (bottom panels) for the VIS2. The free parameter $x$ takes a value of $\sim 3.2\mathrm{e}{-1}$. Figure~\ref{fig:t3fit_SC} shows the same for the T3 and the free parameter $y$ takes a value of $\sim 7.4\mathrm{e}{-2}$.

\subsection{Simulated and real data}
\label{sec:simulated_and_real_data}

\begin{table*}
\caption{GRAVITY data used for the companion injection and recovery tests.}
\label{tab:gravity_data}
\centering
\begin{tabular}{l l l l}
\hline\hline
Programme & Filename & OB date & OB time (UT) \\
\hline
\multirow{3}{*}{60.A-9801(U)} & GRAVI.2019-03-29T01-42-55.145\_singlecalvis.fits & 2019-03-29 & 01:42:53 \\
& GRAVI.2019-03-29T01-51-13.167\_singlecalvis.fits & 2019-03-29 & 01:51:09 \\
& GRAVI.2019-03-29T02-01-37.193\_singlecalvis.fits & 2019-03-29 & 02:01:35 \\
\hline
\multirow{3}{*}{0101.C-0907(B)} & GRAVI.2018-04-18T08-08-19.739\_singlescip2vmred.fits & 2018-04-18 & 08:08:16 \\
& GRAVI.2018-04-18T08-12-10.749\_singlescip2vmred.fits & 2018-04-18 & 08:12:08 \\
& GRAVI.2018-04-18T08-20-04.769\_singlescip2vmred.fits & 2018-04-18 & 08:20:02 \\
\hline
\end{tabular}
\end{table*}

In order to demonstrate the improvement that comes from taking into account the correlations between the data we perform companion injection and recovery tests with simulated and real data. Therefore, we use GRAVITY data of $\zeta$~Ant~B from the technical time programme 60.A-9801(U) and of HIP~78183 from the normal programme 0101.C-0907(B), PI M.~J.~Ireland, listed in Table~\ref{tab:gravity_data}. Both objects were observed in single-field medium resolution mode, but the former one with the four UTs and the latter one with the four ATs (medium configuration D0-G2-J3-K0).

\begin{figure}
\includegraphics[width=\columnwidth]{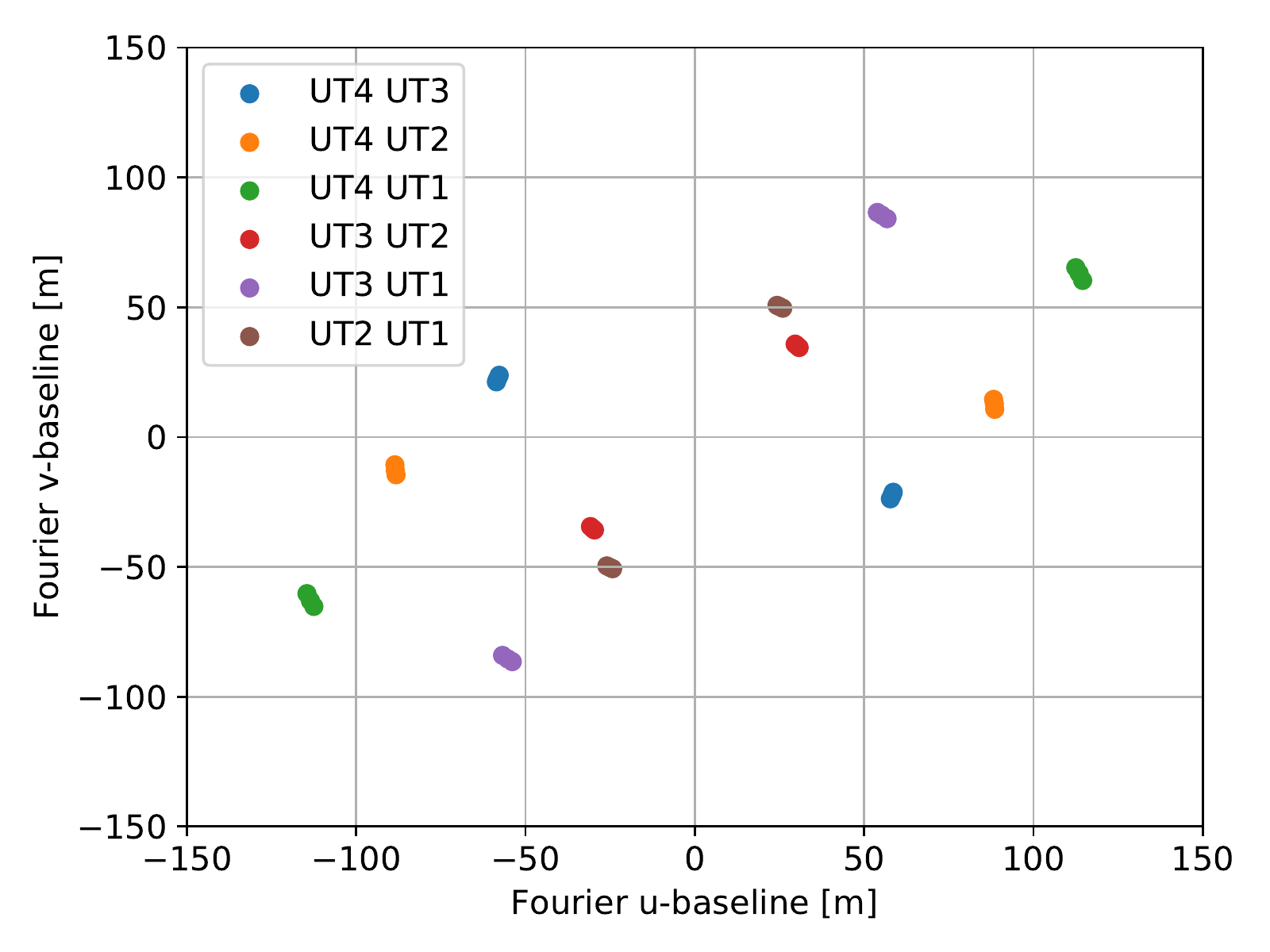}
\caption{Fourier u- and v-baselines of our simulated data extracted from three on-sky observations of GRAVITY using the four UTs over $\sim 20~\text{min}$.}
\label{fig:uv_coverage}
\end{figure}

From the file GRAVI.2019-03-29T02-01-37.193\_singlecalp2vmred.fits we have already extracted the covariances and correlations and fitted our empirical models to them (cf. Section~\ref{sec:empirical_model_for_the_correlations}). For the companion injection and recovery tests with simulated data, we simply use these models and the uv-tracks $u$ and $v$ of the files belonging to programme 60.A-9801(U) listed in Table~\ref{tab:gravity_data} in order to obtain a realistic uv-coverage over $\sim 20~\text{min}$ (cf. Figure~\ref{fig:uv_coverage}). We simulate the complex visibility of a uniform disc with an unresolved companion according to
\begin{align}
    \label{eqn:vis_bin}
    \text{VIS}_\text{bin} &= \frac{\text{VIS}_\text{ud}+f\exp\left(-2\pi i\left(\frac{\Delta_\text{RA}u}{\lambda}+\frac{\Delta_\text{DEC}v}{\lambda}\right)\right)}{1+f}, \\
    \label{eqn:vis_ud}
    \text{VIS}_\text{ud} &= \frac{2J_1(\pi\theta b)}{\pi\theta b},
\end{align}
where $0 \leq f \leq 1$ is the relative flux of the companion, $\Delta_\text{RA}$ and $\Delta_\text{DEC}$ are the on-sky separation in the direction of the celestial north and the celestial east between the companion and its host star, $\lambda$ is the observing wavelength, $J_1$ is the first order Bessel function of first kind, $\theta$ is the angular diameter of the uniform disc and $b = \sqrt{u^2+v^2}$ is the distance between the two telescopes observing the object. The squared visibility amplitudes and the closure phases follow according to
\begin{align}
    \text{VIS2}_\text{bin} &= |\text{VIS}_\text{bin}|^2, \\
    \text{T3}_\text{bin} &= \bm{k} \cdot \angle\text{VIS}_\text{bin}.
\end{align}
Then, we add correlated noise to the $\text{VIS2}_\text{bin}$ and the $\text{T3}_\text{bin}$ by drawing from a multivariate normal distribution with covariance $\bm{\Sigma}_\text{VIS2,fit}$ and $\bm{\Sigma}_\text{T3,fit}$, which we obtain from our correlation model $\bm{C}_\text{VIS2,fit}$ and $\bm{C}_\text{T3,fit}$ (cf. Section~\ref{sec:empirical_model_for_the_correlations}) and assuming a standard deviation of $\sigma_\text{VIS2} = 0.01$ and $\sigma_\text{T3} = 1~\text{deg}$.

For the companion injection and recovery tests with real data, we extract the correlations of the visibility amplitudes VISAMP (instead of the squared visibility amplitudes VIS2) and the closure phases T3 from the P2VM-reduced files belonging to programme 0101.C-0907(B) listed in Table~\ref{tab:gravity_data}, fit our empirical models to them and compute the covariances using Equation~\ref{eqn:corr2cov} and the errors from the corresponding final GRAVITY pipeline products (the ``singlesciviscalibrated'' files). Using the VISAMP instead of the VIS2 can yield better results in some cases where the normalisation of the VIS2 is not done properly by the GRAVITY data reduction pipeline. From the final GRAVITY pipeline products, we also extract the VISAMP and the T3 and inject an unresolved companion according to
\begin{align}
    \label{eqn:inj_visamp}
    \text{VISAMP}_\text{inj} &= \text{VISAMP} \cdot |\text{VIS}_\text{bin}|, \\
    \label{eqn:inj_t3}
    \text{T3}_\text{inj} &= \text{T3}+\bm{k} \cdot \angle\text{VIS}_\text{bin},
\end{align}
where we set $\text{VIS}_\text{ud}$ to one. The VISAMP are simply the square root of the VIS2, so that our correlation model and fitting routines can be equivalently applied in the high SNR regime.

\subsection{Model fitting}
\label{sec:model_fitting}

We search for faint companions in the data by fitting the model for a uniform disc with an unresolved companion (cf. Equation~\ref{eqn:vis_bin}) to it. We maximise the log-likelihood of the model by minimising its $\chi^2$ in order to find the best fit parameters $p_\text{fit}$ of the model, that is
\begin{equation}
    \label{eqn:pfit}
    p_\text{fit} = \mathrm{argmin}_p(\chi^2) = \mathrm{argmin}_p(R^T \cdot \bm{\Sigma}^{-1} \cdot R),
\end{equation}
where $R = D-M$ is the residual between data and model and $p = (f, \Delta_\text{RA}, \Delta_\text{DEC}, \theta)$ is the four-dimensional parameter vector of the model.

In order to find the global minimum of the $\chi^2$ within a given range of companion separations, we first find a prior for the uniform disc diameter $\theta_0$ by fitting the corresponding model (cf. Equation~\ref{eqn:vis_ud}) to the data. Then, we perform a set of minimisations with priors on a $\Delta_\text{RA}$-$\Delta_\text{DEC}$ grid, the uniform disc diameter $\theta_0$, and a small relative flux $f_0 = 1\mathrm{e}{-3}$. This is necessary since the $\chi^2$ hyper-surface is bumpy (i.e. has many local extrema) if projected onto the $\Delta_\text{RA}$-$\Delta_\text{DEC}$ surface and the BFGS algorithm which is used to minimise the $\chi^2$ converges on local minima. The bumpiness is a result of the sparse uv-coverage of a long-baseline optical interferometer which causes the sensitivity to vary substantially over the FOV.

The above method relies on the covariance matrix $\bm{\Sigma}$ being invertible. This is not the case for a sample covariance that is estimated from a small number of frames, which is usually singular, and is the reason why we develop an empirical covariance model. However, our empirical model for the covariances of the closure phases is also singular, since the fourth triangle can be written as a linear combination of the other three. There are multiple solutions to this problem, and for simplicity we decide to completely ignore the data recorded on the fourth triangle since it theoretically is redundant anyway\footnote{In practice, this is not the case since the data is affected by different errors originating from different optical paths through the instrument and different detector noise.}. There are more sophisticated methods to keep the data recorded on the fourth triangle, such as the ``jackknife'' method (i.e. averaging over four model fits using data recorded on different sets of three triangles), projection into a sub-space that preserves the information in the covariance matrix \citep{blackburn2020}, and the approach from \citet{kulkarni1989} which is adding a small numerical value $\epsilon \ll 1$ to the diagonal of the covariances of the closure phases, that is
\begin{equation}
    \bm{\Sigma}_\text{T3,invertible} = \bm{\Sigma}_\text{T3,fit}+\epsilon \cdot \mathrm{id},
\end{equation}
where $\mathrm{id}$ is the identity matrix, so that the covariance matrix becomes numerically invertible.

Finally, in order to determine the statistical significance of a detected companion, we compute the probability that the binary model is preferred over the uniform disc model according to
\begin{equation}
    \label{eqn:detection_criterion}
    P = 1-\text{CDF}_{N_\text{dof}}\left(\frac{N_\text{dof}\chi^2_\text{red,ud}}{\chi^2_\text{red,bin}}\right),
\end{equation}
where $\text{CDF}_{N_\text{dof}}$ is the $\chi^2$ cumulative distribution function with $N_\text{dof}$ degrees of freedom, $\chi^2_\text{red,ud}$ is the reduced $\chi^2$ of the best fit uniform disc model and $\chi^2_\text{red,bin}$ is the reduced $\chi^2$ of the best fit binary model \citep[cf.][]{gallenne2015}.

If the host star is essentially unresolved (i.e. $\theta b\lambda \ll 1$) and the companion is at high contrast (i.e. $f \ll 1$) one can linearise the $\text{VIS2}_\text{bin}$ and the $\text{T3}_\text{bin}$ as a function of the relative flux of the companion $f$ according to
\begin{align}
    \text{VIS2}_\text{bin} &\propto 1+f, \\
    \text{T3}_\text{bin} &\propto f.
\end{align}
A more detailed derivation of this relationship can be found in Appendix~\ref{app:linearised_model}. Let $D$ be the data, $M_\text{ref}$ a reference binary model which is normalised to the relative flux of the companion $f_\text{ref}$, and $\bm{\Sigma}$ the covariances between the data, that is
\begin{align}
    D &= \begin{pmatrix}
    \text{VIS2}-1 \\
    \text{T3}
    \end{pmatrix}, \\
    M_\text{ref} &= \begin{pmatrix}
    (\text{VIS2}_\text{bin,ref}-1)/f_\text{ref} \\
    \text{T3}_\text{bin,ref}/f_\text{ref}
    \end{pmatrix}, \\
    \bm{\Sigma} &= \begin{pmatrix}
    \bm{\Sigma}_\text{VIS2} & \bm{0} \\
    \bm{0} & \bm{\Sigma}_\text{T3}
    \end{pmatrix},
\end{align}
where $\text{VIS2}_\text{bin,ref}$ and $\text{T3}_\text{bin,ref}$ are the binary model VIS2 and T3 evaluated at a reference relative flux $f_\text{ref} = 1\mathrm{e}{-3}$. Then, the best fit relative flux $f_\text{fit}$ and its uncertainty $\sigma_{f_\text{fit}}$ follow according to
\begin{align}
    \label{eqn:ffit}
    f_\text{fit} &= \frac{M_\text{ref}^T \cdot \bm{\Sigma}^{-1} \cdot D}{M_\text{ref}^T \cdot \bm{\Sigma}^{-1} \cdot M_\text{ref}}, \\
    \sigma_{f_\text{fit}} &= \frac{1}{\sqrt{M_\text{ref}^T \cdot \bm{\Sigma}^{-1} \cdot M_\text{ref}}},
\end{align}
\citep[cf.][]{lebouqin2012,kammerer2019}. Equation~\ref{eqn:ffit} can be computed on a $\Delta_\text{RA}$-$\Delta_\text{DEC}$ grid, and the best fit parameters $p_\text{fit}$ follow from the grid position which minimises
\begin{equation}
    \chi^2_\text{red} = \frac{\chi^2}{N_\text{dof}} = \frac{R^T \cdot \bm{\Sigma}^{-1} \cdot R}{N_\text{dof}},
\end{equation}
where $N_\text{dof}$ is the number of the degrees of freedom. This grid search technique is commonly used in order to find the global minimum of the $(f, \Delta_\text{RA}, \Delta_\text{DEC}, \theta)$ parameter space and its corresponding $\chi^2_\text{red}$ \citep[e.g.][]{absil2011,gallenne2015}. However, the statistical structure of the grid is complex due to redundancy and periodicity in sensitivity originating from the very limited uv-coverage of a sparse interferometer such as the VLTI (cf. Figure~\ref{fig:uv_coverage}). Therefore, the detection significance is derived using Equation~\ref{eqn:detection_criterion} which yields the probability that the binary model is preferred over the uniform disc model (without any companion).


\section{Results}
\label{sec:results}

We evaluate the impact of our full covariance model by performing model fitting and companion injection and recovery tests with both simulated and real GRAVITY data. In Section~\ref{sec:model_fitting_to_correlated_noise}, we simulate data without any astronomical object, that is correlated noise only, and use model fitting to determine the fundamental detection limits when assuming uncorrelated data (i.e. a diagonal covariance) and correlated data (i.e. our full covariance model). In Sections~\ref{sec:injection_and_recovery_tests_simulated_data} and~\ref{sec:injection_and_recovery_tests_real_data} we inject companions with different relative fluxes and separations into simulated and real GRAVITY data and try to recover them, again assuming both uncorrelated and correlated data.

\subsection{Model fitting to correlated noise}
\label{sec:model_fitting_to_correlated_noise}

In order to compare the fundamental detection limits when assuming uncorrelated and correlated data, we simulate 100 GRAVITY data sets of an unresolved host star without any companion (i.e. $\theta = f = 0$ in Equation~\ref{eqn:vis_bin}) affected by realistic correlated noise ($\bm{\Sigma}_\text{VIS2,fit}$ and $\bm{\Sigma}_\text{T3,fit}$, cf. Section~\ref{sec:simulated_and_real_data}). Then, we use Equation~\ref{eqn:ffit} in order to compute the best fit relative flux $f_\text{fit}$ on a $\Delta_\text{RA}$-$\Delta_\text{DEC}$ grid for each of the 100 simulated data sets, first assuming uncorrelated data (i.e. a diagonal covariance $\mathrm{diag}(\bm{\Sigma}_\text{VIS2,fit})$ and $\mathrm{diag}(\bm{\Sigma}_\text{T3,fit})$) and then assuming correlated data (i.e. our full covariance model) in Equation~\ref{eqn:ffit}. Since no companion was injected into the data, the best fit flux ratios of these grids represent the fundamental contrast floor. Any companion with a higher contrast (i.e. smaller flux) would not be distinguishable from the noise. By computing an azimuthal average of these grids, we obtain a fundamental 1--$\sigma$ contrast curve (i.e. best fit relative flux vs. angular separation curve).

\begin{figure*}
\includegraphics[width=\columnwidth]{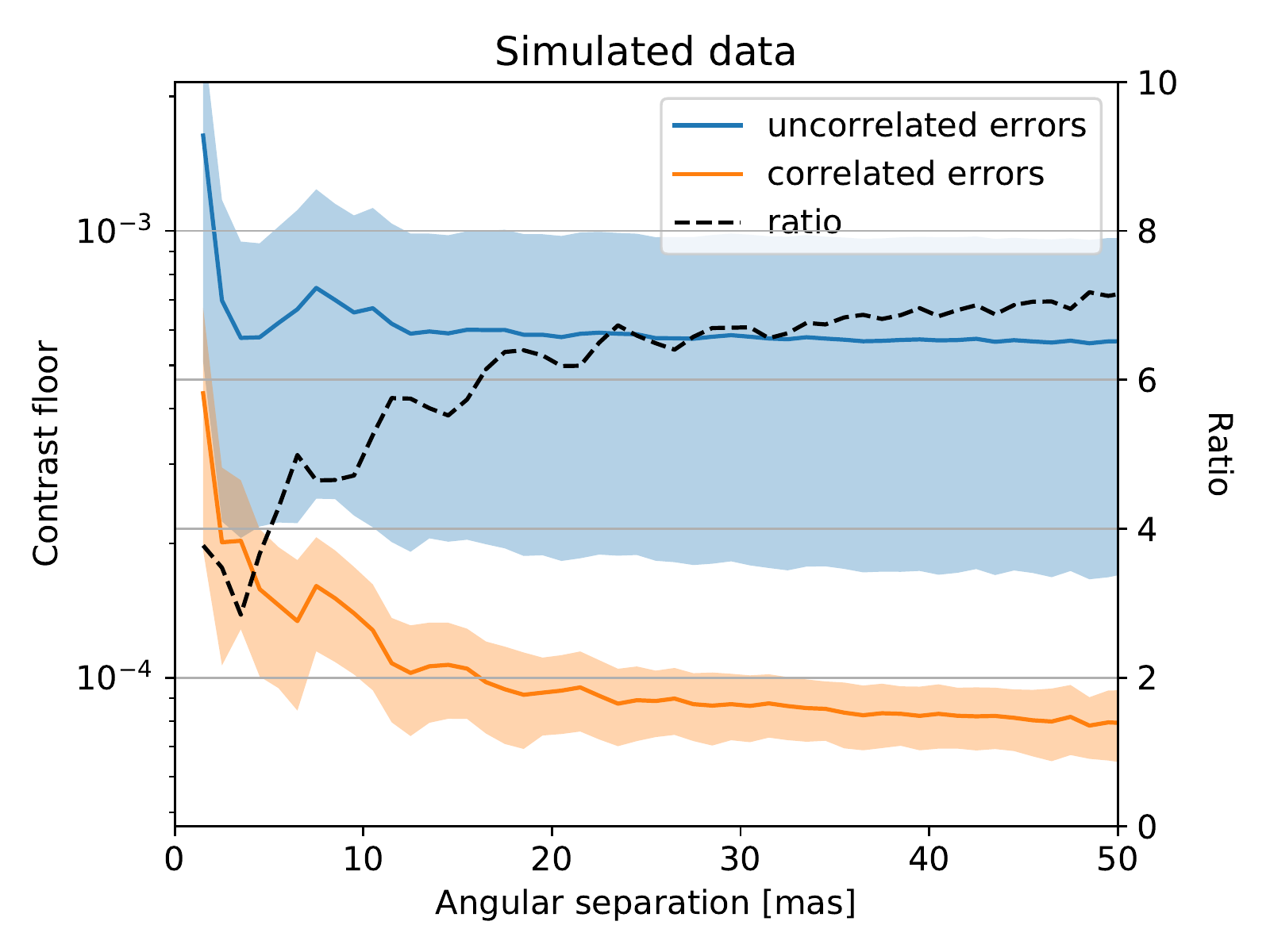}
\includegraphics[width=\columnwidth]{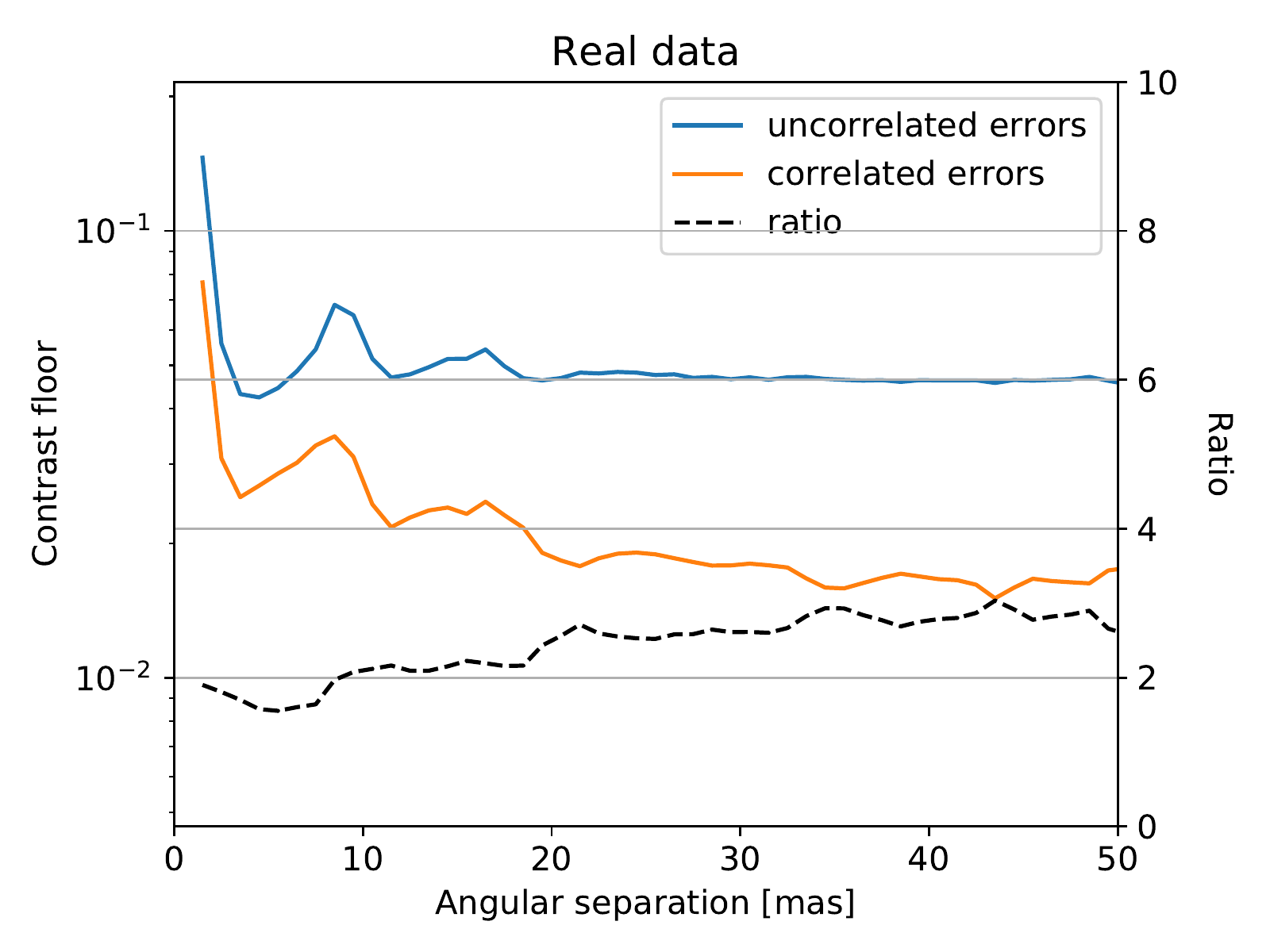}
\caption{Left panel: contrast curve (i.e. azimuthal average of the best fit relative flux) for simulated data of an unresolved host star without any companion affected by correlated errors, computed with model fitting assuming uncorrelated data (blue curve) and correlated data (orange curve). Both curves show the mean contrast curve over 100 simulated data sets and the shaded region highlights its standard deviation. The dashed black line shows the ratio of the blue and the orange curve, representing the improvement (i.e. the factor by which the detection limits improve) when using our correlated error model instead of the classical uncorrelated one}. Right panel: same, but for the real GRAVITY data introduced in Section~\ref{sec:simulated_and_real_data}.
\label{fig:fit_to_noise}
\end{figure*}

The mean of the 100 azimuthal averages obtained for each of these two scenarios (model fitting assuming uncorrelated data in blue and correlated data in orange) is shown in the left panel of Figure~\ref{fig:fit_to_noise}. The contrast floor remains roughly constant at a contrast of $\sim 6\mathrm{e}{-4}$ outward of an angular separation of $\sim 5~\text{mas}$ for the scenario assuming uncorrelated data. This is because at a contrast of $\sim 6\mathrm{e}{-4}$ one is dominated by the systematic (i.e. the correlated) errors. However, for the scenario assuming correlated data, the fundamental 1--$\sigma$ detection limit continues to decrease with increasing angular separation. At an angular separation of $\sim 10~\text{mas}$ it is already a factor of four better than the limit assuming uncorrelated data (cf. dashed black curve). We note that such a behaviour has already been observed by \citet{ireland2013} for orthogonal kernel phases and statistically independent kernel phases, which are obtained by projecting the orthogonal kernel phases into an eigenspace with zero covariances. Its reason is that at small angular separations, the detection limits rely on the average of the VIS2 and the T3 over the spectral channels, while with increasing angular separations the VIS2 and T3 vary within the spectral bands and the impact of the correlations is growing. A flat uncorrelated contrast curve (as a function of angular separation) is further consistent with previous works on interferometric observables assuming uncorrelated data \citep[e.g.][]{absil2011,gallenne2015}.

Furthermore, when using our full covariance model, the contrast floor is also more stable for different representations of the noise (highlighted by the shaded regions in the left panel of Figure~\ref{fig:fit_to_noise} which show the standard deviation of the contrast curves over the 100 simulated data sets) meaning that the derived detection limits can be regarded more robust (i.e. independent of the exact representation of the noise which is a random component). Hence, if an observer is only working with a small number of data sets, they will still be able to derive universally valid detection limits.

The right panel of Figure~\ref{fig:fit_to_noise} shows the same plot, but for model fitting to the real GRAVITY data consisting of the three files belonging to programme 0101.C-0907(B) listed in Table~\ref{tab:gravity_data}. Since there is only one real GRAVITY data set we cannot compute or show any standard deviation. The plot looks similar, except for the fundamental detection limits being about two orders of magnitude worse and the ratio between the two scenarios being a lot smaller due to much weaker correlations being present in the real data if compared to the simulated data (cf. Figure~\ref{fig:z_corr_SC}).

\subsection{Injection and recovery tests (simulated data)}
\label{sec:injection_and_recovery_tests_simulated_data}

As a next step, we perform companion injection and recovery tests with simulated data, in order to compare the empirical detection limits when assuming uncorrelated and correlated data. Therefore, we simulate GRAVITY data sets (affected by correlated noise) of a $1~\text{mas}$ uniform disc (the host star) and inject companions with a range of relative fluxes and at different positions around the host star, that is
\begin{align}
    f &\in [10^{-4},10^{-3.75},10^{-3.5},...,10^{-1.5}], \\
    \Delta_\text{RA} &\in [-30,-25,-20,...,30]~\text{mas}, \\
    \Delta_\text{DEC} &\in [-30,-25,-20,...,30]~\text{mas},
\end{align}
using Equation~\ref{eqn:vis_bin}. Then, we perform model fitting with priors on a $\Delta_\text{RA}$-$\Delta_\text{DEC}$ grid in order to find the global minimum of the $\chi^2$ (cf. Equation~\ref{eqn:pfit}). We note that this method is similar to how CANDID searches for companions for example. Similar to before (cf. Section~\ref{sec:model_fitting_to_correlated_noise}), we perform the model fitting once assuming uncorrelated data (i.e. a diagonal covariance $\mathrm{diag}(\bm{\Sigma}_\text{VIS2,fit})$ and $\mathrm{diag}(\bm{\Sigma}_\text{T3,fit})$) and once assuming correlated data (i.e. our full covariance model). We classify an injected companion as recovered if the best fit relative flux $f_\text{fit}$ differs by no more than 10\% from the injected one $f_\text{inj}$ and the best fit position $(\Delta_\text{RA,fit},\Delta_\text{DEC,fit})$ differs by no more than one resolution element of the interferometer from the injected one $(\Delta_\text{RA,inj},\Delta_\text{DEC,inj})$, that is
\begin{align}
    |f_\text{fit}-f_\text{inj}|/f_\text{inj} &< 0.1, \\
    \sqrt{(\Delta_\text{RA,fit}-\Delta_\text{RA,inj})^2+(\Delta_\text{DEC,fit}-\Delta_\text{DEC,inj})^2} &< \frac{\lambda_\text{mean}}{2b_\text{max}},
\end{align}
where $\lambda_\text{mean}$ is the mean of the observed wavelength range ($\sim 2.2~\text{\textmu m}$ for GRAVITY) and $b_\text{max}$ is the longest baseline of the interferometer ($\sim 130~\text{m}$ for observations with the VLTI UTs).

\begin{figure*}
\includegraphics[width=\columnwidth]{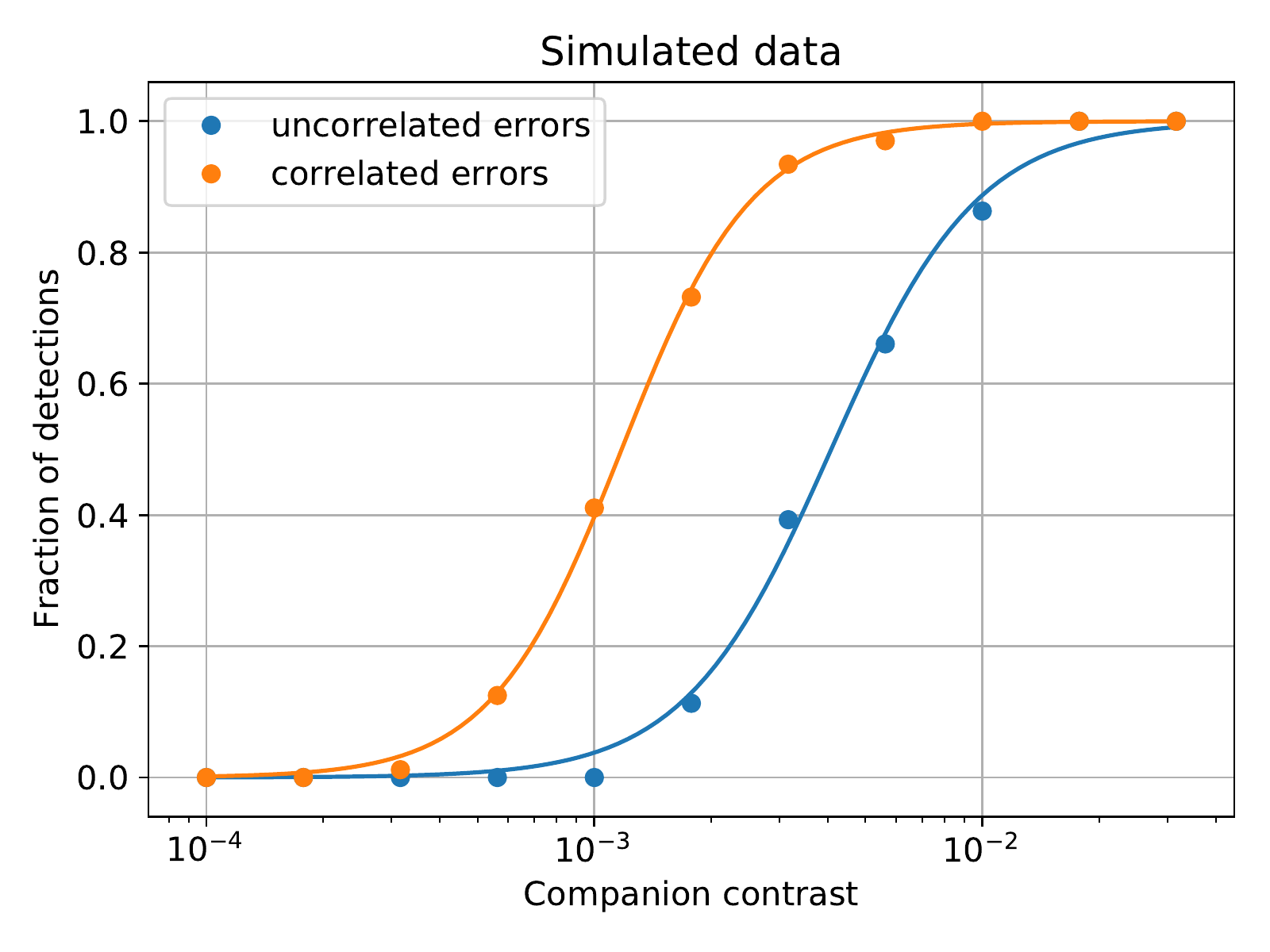}
\includegraphics[width=\columnwidth]{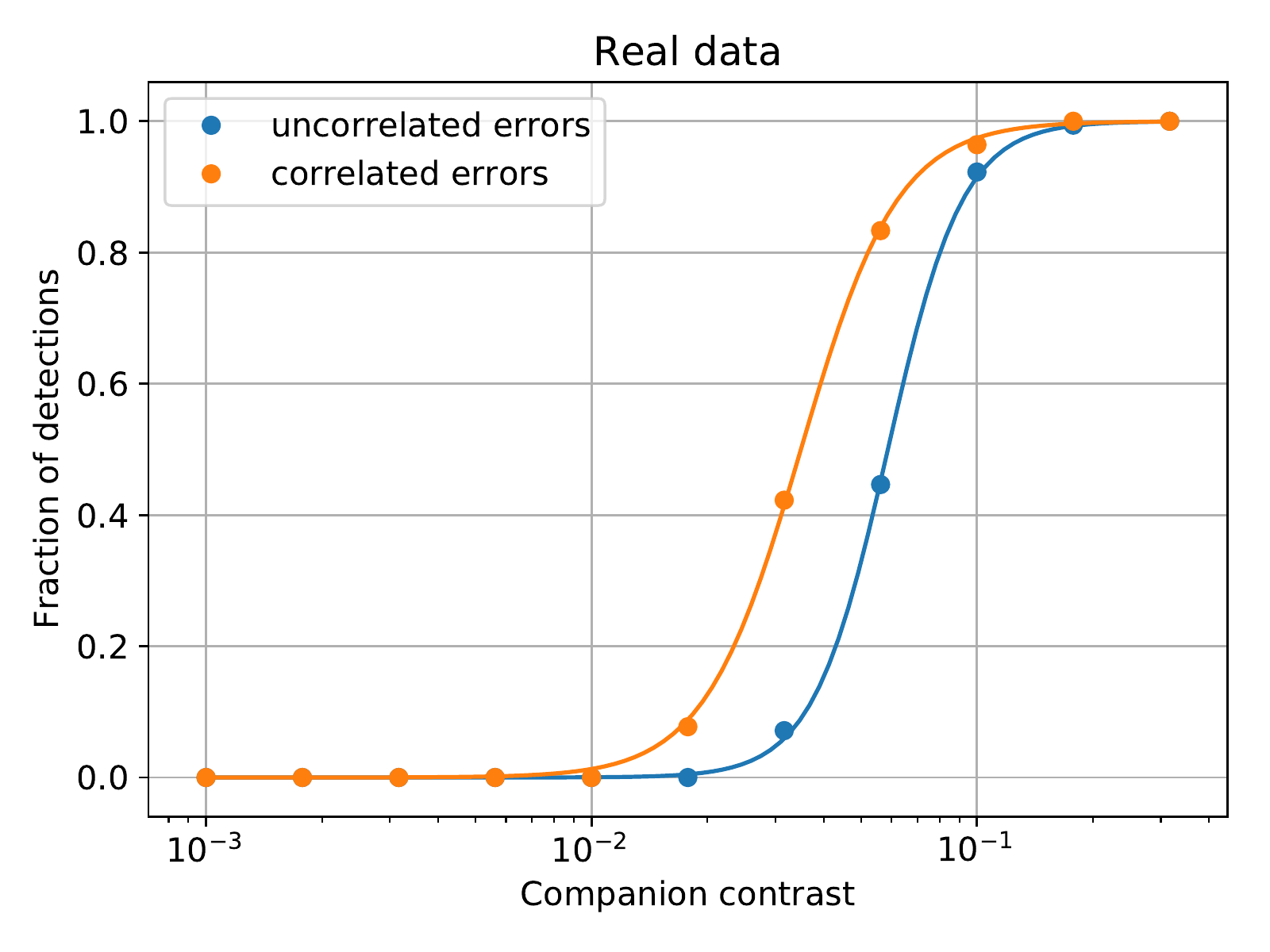}
\caption{Left panel: fraction of correctly recovered companions as a function of the relative flux of the injected companion from injection and recovery tests with simulated GRAVITY data (cf. Section~\ref{sec:injection_and_recovery_tests_simulated_data}), assuming uncorrelated data (blue points) and correlated data (orange points) for the model fitting. The blue and the orange curves are logistic growth functions fitted to the data points. Right panel: same, but from injection and recovery tests with real GRAVITY data (cf. Section~\ref{sec:injection_and_recovery_tests_real_data}).}
\label{fig:injection_recovery}
\end{figure*}

The left panel of Figure~\ref{fig:injection_recovery} shows the fraction of recovered companions as a function of the relative flux of the companion for model fitting assuming uncorrelated data (blue points) and correlated data (orange points). These values are summed over all positions around the host star with $5~\text{mas} \leq \rho \leq 45~\text{mas}$, where $\rho = \sqrt{\Delta^2_\text{RA,inj}+\Delta^2_\text{DEC,inj}}$ is the angular separation, so to avoid any significant influence from the $1~\text{mas}$ uniform disc (the host star). Although one of the findings in this paper is that the contrast curve is not flat outward a few $\lambda/b_\text{max}$ when accounting for the data correlations (cf. Figure~\ref{fig:fit_to_noise}), it is still a reasonable simplification to sum over positions with different angular separations.

The overplotted blue and orange curves are logistic growth functions
\begin{equation}
    g(x) = \frac{L}{1+e^{-k(x-x_0)}},
\end{equation}
fitted to the data points (in log-space), where $L = 1$ is the upper growth barrier, $k$ is the logistic growth rate and $x_0$ is the midpoint. The fact that the orange curve is shifted towards the left compared to the blue curve means that the detection limits are fainter when assuming correlated data instead of uncorrelated data. This could be expected since the data is affected by correlated noise and correctly accounting for these correlations in the model fitting should lead to fainter detection limits. For any given fraction of detections (i.e. any point on the y-axis), the ratio of the contrasts (i.e. the x-values) of the blue and the orange curve gives the improvement that comes from our full covariance model over the conventional diagonal covariance model. This ratio varies between $\sim 3$ and $\sim 4$, depending on the fraction of detections, with an average value of $\sim 3.5$. This means that the detection limits improve by a factor of $\sim 3.5$ when assuming correlated data instead of uncorrelated data.

In order to check the validity of our detection criterion (cf. Equation~\ref{eqn:detection_criterion}) we count the number of companions in different categories, which we represent in a confusion matrix
\begin{equation}
    \zeta = \begin{pmatrix}
    \text{\# of true positives} & \text{\# of false negatives} \\
    \text{\# of false positives} & \text{\# of true negatives}
    \end{pmatrix},
\end{equation}
where positive/negative refers to a detection being classified as significant/insignificant according to our detection criterion. Hence, in an ideal world, there would only be true positives or true negatives and $\zeta$ would be a diagonal matrix. Obviously, this is not the case in the real world where the data is affected by noise. If we choose an optimistic detection criterion (e.g. the significance needs to be above 1--$\sigma$) there will be many false positives (i.e. many detections that are classified as significant, but which are no true companions) and if we choose a pessimistic detection criterion (e.g. the significance needs to be above 5--$\sigma$) there will be many false negatives (i.e. many detections that are classified as insignificant, but which are true companions).

The confusion matrices from our companion injection and recovery tests for a 3--$\sigma$ detection criterion are
\begin{align}
    \zeta_\text{diag} &= \begin{pmatrix}
    674 & 3 \\
    517 & 654
    \end{pmatrix}, \\
    \zeta_\text{full} &= \begin{pmatrix}
    826 & 213 \\
    12 & 797
    \end{pmatrix},
\end{align}
for assuming uncorrelated data ($\zeta_\text{diag}$) and correlated data ($\zeta_\text{full}$). In the former case, there is a large fraction of false detections being classified as significant ($517/1171 \approx 44\%$), whereas in the latter case this fraction ($12/809 \approx 1\%$) is roughly consistent with a 3--$\sigma$ result. The number of true detections is higher when assuming correlated data (1039 = 826+213) than when assuming uncorrelated data (677 = 674+3) because the detection limits are fainter, and although the number of false negatives (i.e. true companions being classified as insignificant) is a lot higher, the number of true positives is still higher when assuming correlated data. In summary, when using our full covariance model there are less detections above 3--$\sigma$ significance than when using the conventional diagonal covariance model (838 = 826+12 vs. 1191 = 674+517), but the number of true positives (i.e. true companions being classified as significant) is still higher and significant detections are much more reliable since there are almost no false positives. Accounting for the correlations is therefore clearly preferred over ignoring them.

Before proceeding to the injection and recovery tests with real data we also assess the robustness of our correlation model with respect to errors in the model parameters $x$ and $y$. Therefore, we repeat the injection and recovery tests with wrong correlation and covariance matrices where $x$ and $y$ are only 50\% and 25\% of their true values respectively. We find that the number of false positives or false negatives increases slightly, but not significantly. This was expected since the detection of asymmetric structure (such as a companion) is governed by the T3 whose correlations are dominated by the correlations of $\pm 1/3$ originating from shared baselines among different triangles. For scenarios where the VIS2 have a larger impact on the model (e.g. measuring stellar diameters) we expect that errors in the model parameters, especially $x$, have a more significant impact.

\subsection{Injection and recovery tests (real data)}
\label{sec:injection_and_recovery_tests_real_data}

In the previous Section it is obvious that our full covariance model would outperform the conventional diagonal covariance model, since we simulated data affected by correlated noise. Therefore, the crucial next step is to validate our methods with real GRAVITY data sets. For this purpose, we extract the correlations of the VISAMP and the T3 from the files belonging to programme 0101.C-0907(B) listed in Table~\ref{tab:gravity_data}, fit our empirical models to them and use the VISAMP and the T3 data from the corresponding final GRAVITY pipeline products as noise model.

Since the real data is affected by bright speckles arising from an imperfect calibration, for which our correlation model does not account, we subtract the theoretical VISAMP and T3 of the best fit companion from the data before performing the injection and recovery tests. This also helps us to enter the medium-contrast regime ($f_\text{fit} \lessapprox 10\%$) where the linearisation of the binary model (cf. Appendix~\ref{app:linearised_model}) holds. The parameters of the subtracted best fit companion are
\begin{equation}
    p_\text{sub} = (f, \Delta_\text{RA}, \Delta_\text{DEC}, \theta) = (0.0383, 0.20, -5.46, 0),
\end{equation}
and the corresponding detection map is shown in Figure~\ref{fig:detection_map}. The parameters were obtained assuming correlated errors. An extension of our correlation model to inter-observation correlations, for instance arising from the calibration process, is left for future work.

Then, we compute the covariances from the correlations, the VISAMPERR, and the T3ERR from the final GRAVITY pipeline products using Equation~\ref{eqn:corr2cov} and inject companions with
\begin{align}
    f &\in [10^{-3},10^{-2.75},10^{-2.5},...,10^{-0.5}], \\
    \Delta_\text{RA} &\in [-30,-25,-20,...,30]~\text{mas}, \\
    \Delta_\text{DEC} &\in [-30,-25,-20,...,30]~\text{mas},
\end{align}
using Equations~\ref{eqn:inj_visamp} and~\ref{eqn:inj_t3}. In order to obtain empirical detection limits when assuming uncorrelated and correlated data, we then repeat the model fitting described in the previous Section.

The fraction of correctly recovered companions as a function of the relative flux of the companion for both scenarios (uncorrelated noise: blue points and correlated noise: orange points) is shown in the right panel of Figure~\ref{fig:injection_recovery}, again overplotted with logistic growth functions fitted to the data points (cf. Section~\ref{sec:injection_and_recovery_tests_simulated_data}). The plot looks similar to the one from the injection and recovery tests with simulated data and confirms the applicability of our full covariance model to real GRAVITY data. Of course, the empirical detection limits are about one to two orders of magnitude worse and the improvement that comes from our full covariance model (i.e. the lateral shift of the orange curve with respect to the blue curve) is only a factor of $\sim 2$ (consistent with the right panel of Figure~\ref{fig:fit_to_noise} which also shows an improvement by a factor of $\sim 2$) due to weaker correlations being present in the data used for the injection and recovery tests with real data. In summary, our full covariance model still brings a singificant improvement over the convential diagonal covariance model.


\section{Conclusions}
\label{sec:conclusions}

Correlated noise is placing fundamental detection limits on interferometric data. From on-sky VLTI/GRAVITY data, we extract and illustrate the correlations present in the data and develop an empirical model in order to describe them. This empirical model is sufficiently simple for it to be fitted to the correlations extracted from a single GRAVITY data product and could therefore be directly integrated into the GRAVITY data reduction pipeline and made available to the community as part of the OIFITS 2 file \citep[which has a well-defined standard for providing covariance matrices][]{duvert2017}.

Then, we evaluate the impact of our full covariance model by performing model fitting and companion injection and recovery tests with both simulated and real GRAVITY data. Our methods are based on $\chi^2 = R^T \cdot \bm{\Sigma}^{-1} \cdot R$ minimisation, where we compare the scenarios assuming uncorrelated data (i.e. a diagonal covariance matrix $\mathrm{diag}(\bm{\Sigma})$) and correlated data (i.e. a full covariance matrix $\bm{\Sigma}$ following from our empirical correlation model). We show that accounting for the correlations that we find to be present in GRAVITY data could yield to an improvement in the detection limits by a factor of up to $\sim 3.5$ over ignoring them. Moreover, the obtained detection limits (and therefore also potential detections) can be regarded more robust in the former case. We also highlight the problems which arise from ignoring the correlations, as it is done in model fitting pipelines such as LITpro \citep{tallon-bosc2008} and CANDID \citep{gallenne2015} so far, and discuss that conventional detection criteria based on $\chi^2$ statistics are strongly biased towards false positives (i.e. detections that are no true companions).

The empirical correlation model presented in this paper is a simple one-parameter model derived from GRAVITY data, but is arguably also applicable (with small modifications) to other instruments such as VLTI/PIONIER for example. It only treats the correlations between the different observables, but not yet between different frames or targets (such as the science and the calibrator target). We choose this approach in order to enable a simple implementation into existing data reduction and model fitting pipelines. Especially with the increasing availability of computing power, the use of full covariance matrices for describing the correlated noise in interferometric data should become a standard. Collaborations around future instruments should provide estimated data covariances as part of the official data reduction pipelines.

In the future, we aim to compare our empirical correlation model with the data covariances derived from bootstrapping \citep[e.g.][]{lachaume2019} and extend our model in order to account for correlations between different frames and targets. Finally, we will re-analyse several marginal detections of companions around Cepheid stars from \citet{gallenne2013,gallenne2014,gallenne2015} by properly accounting for the correlated noise.

\begin{acknowledgements}
The authors would like to thank Jean-Baptiste Le Bouquin for helpful feedback on the GRAVITY data reduction pipeline. MJI acknowledges support from the ESO visitor programme. This work has been partly supported by the Australian Research Council's Discovery Projects (DP190101477). The manuscript was also substantially improved following helpful comments from an anonymous referee.
\end{acknowledgements}


\begin{appendix}

\section{Collection of correlations}
\label{app:collection_of_correlations}

\begin{figure*}
\begin{center}
\includegraphics[width=0.8\textwidth]{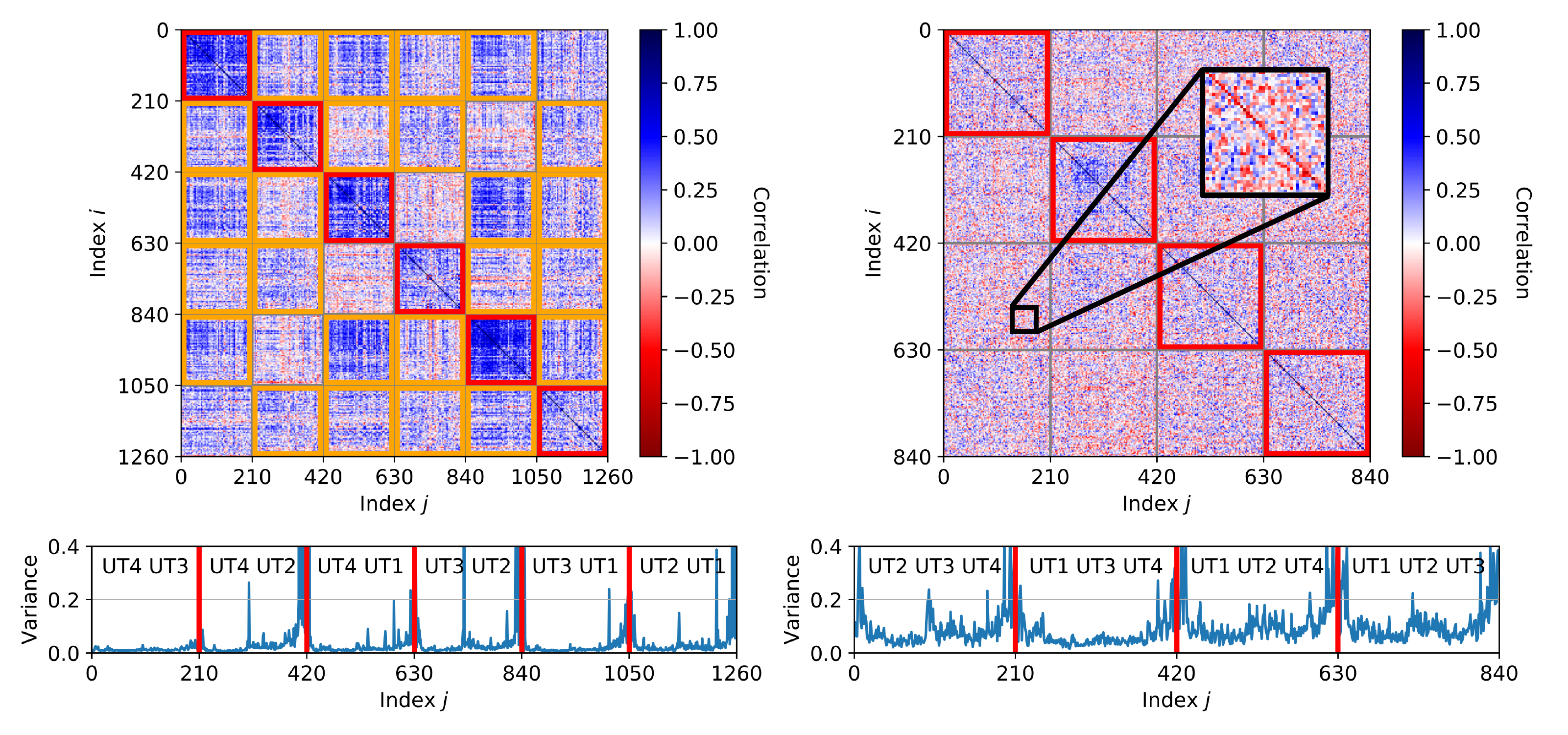}
\includegraphics[width=0.8\textwidth]{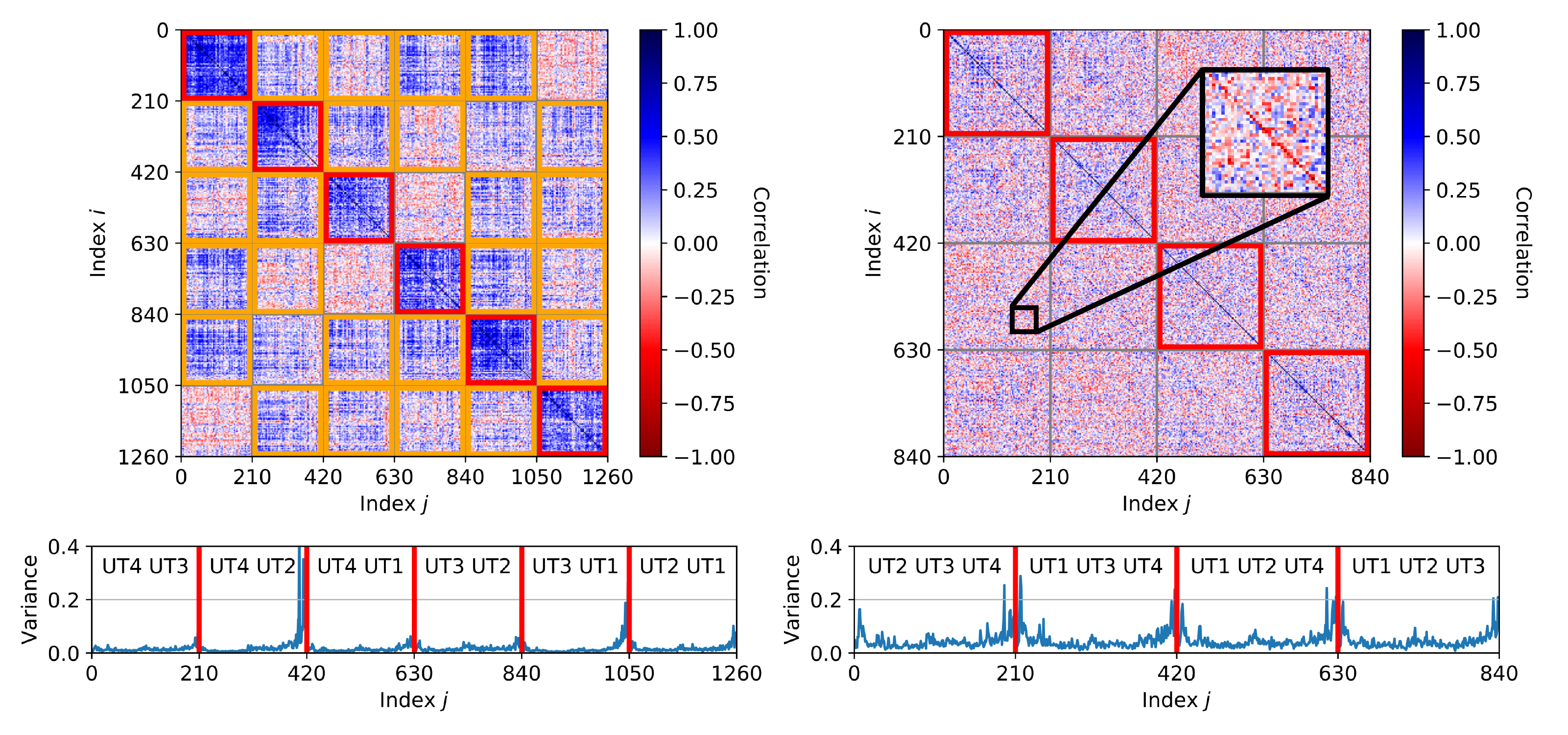}
\includegraphics[width=0.8\textwidth]{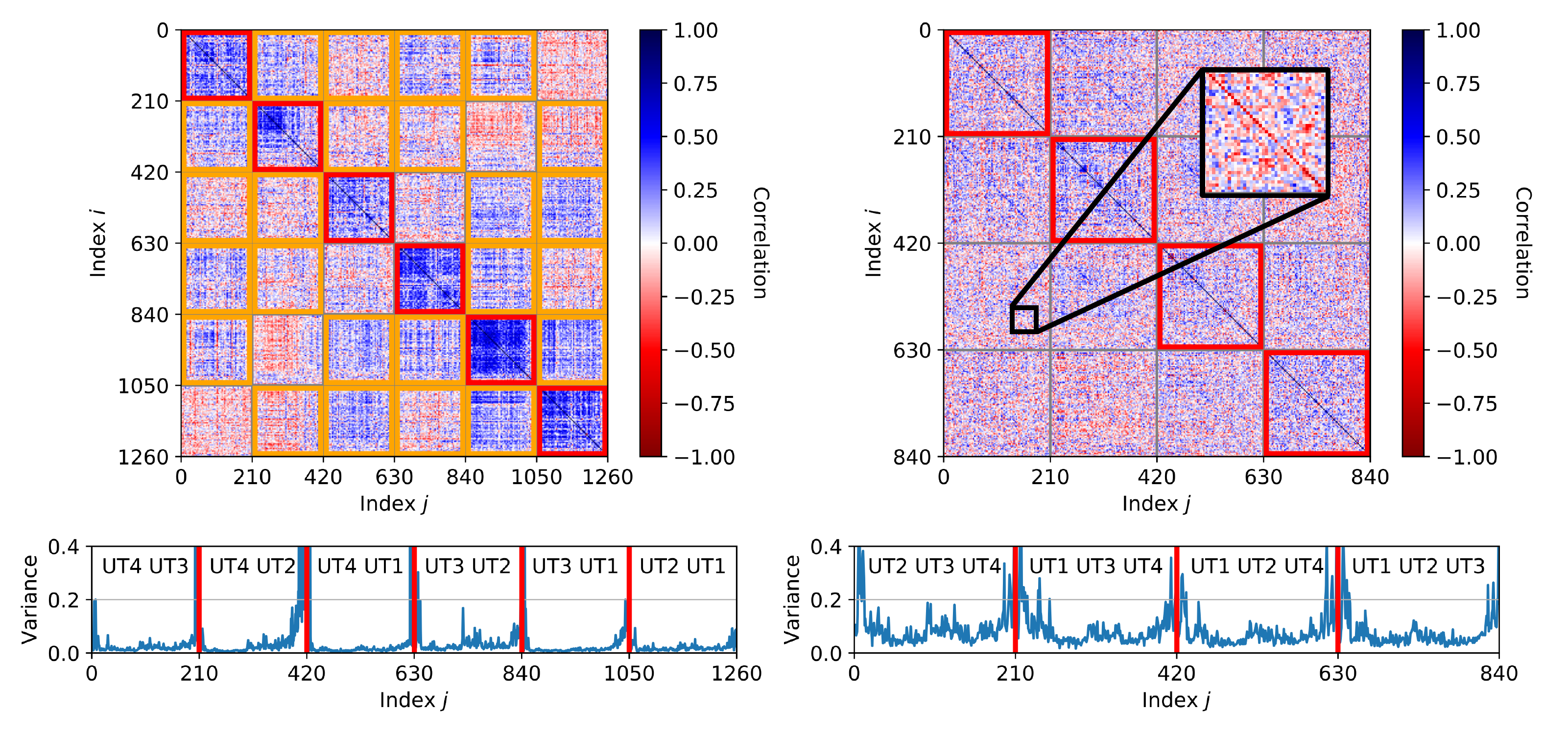}
\end{center}
\caption{Same as Figure~\ref{fig:corr_SC}, but extracted from three other P2VM-reduced files from programme 60.A-9801(U). From top to bottom: GRAVI.2019-03-29T01-46-28.155\_singlecalp2vmred.fits, GRAVI.2019-03-29T01-57-13.182\_singlecalp2vmred.fits, GRAVI.2019-03-29T01-59-31.188\_singlecalp2vmred.fits.}
\label{fig:z_corr_SC}
\end{figure*}

\begin{figure*}
\begin{center}
\includegraphics[width=0.8\textwidth]{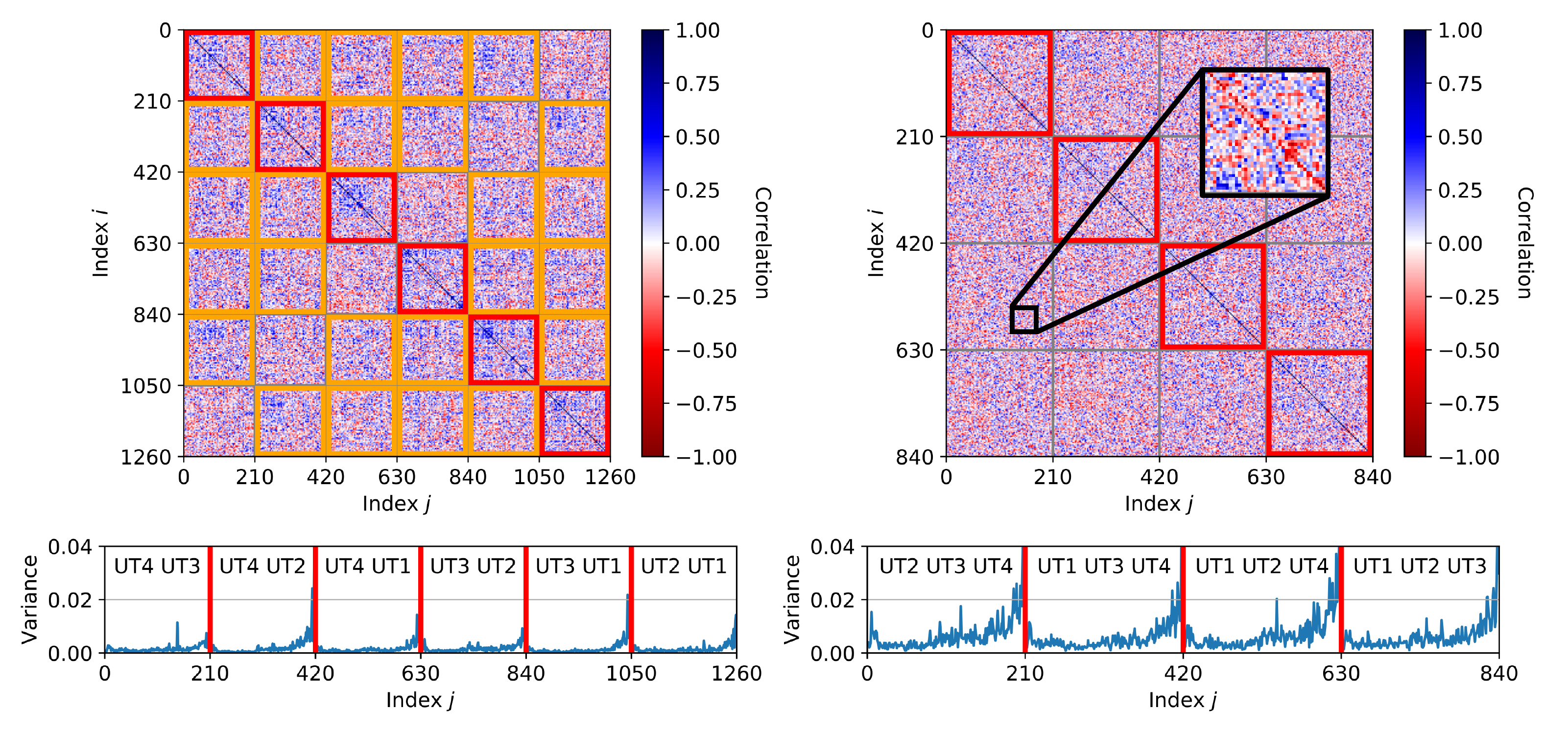}
\includegraphics[width=0.8\textwidth]{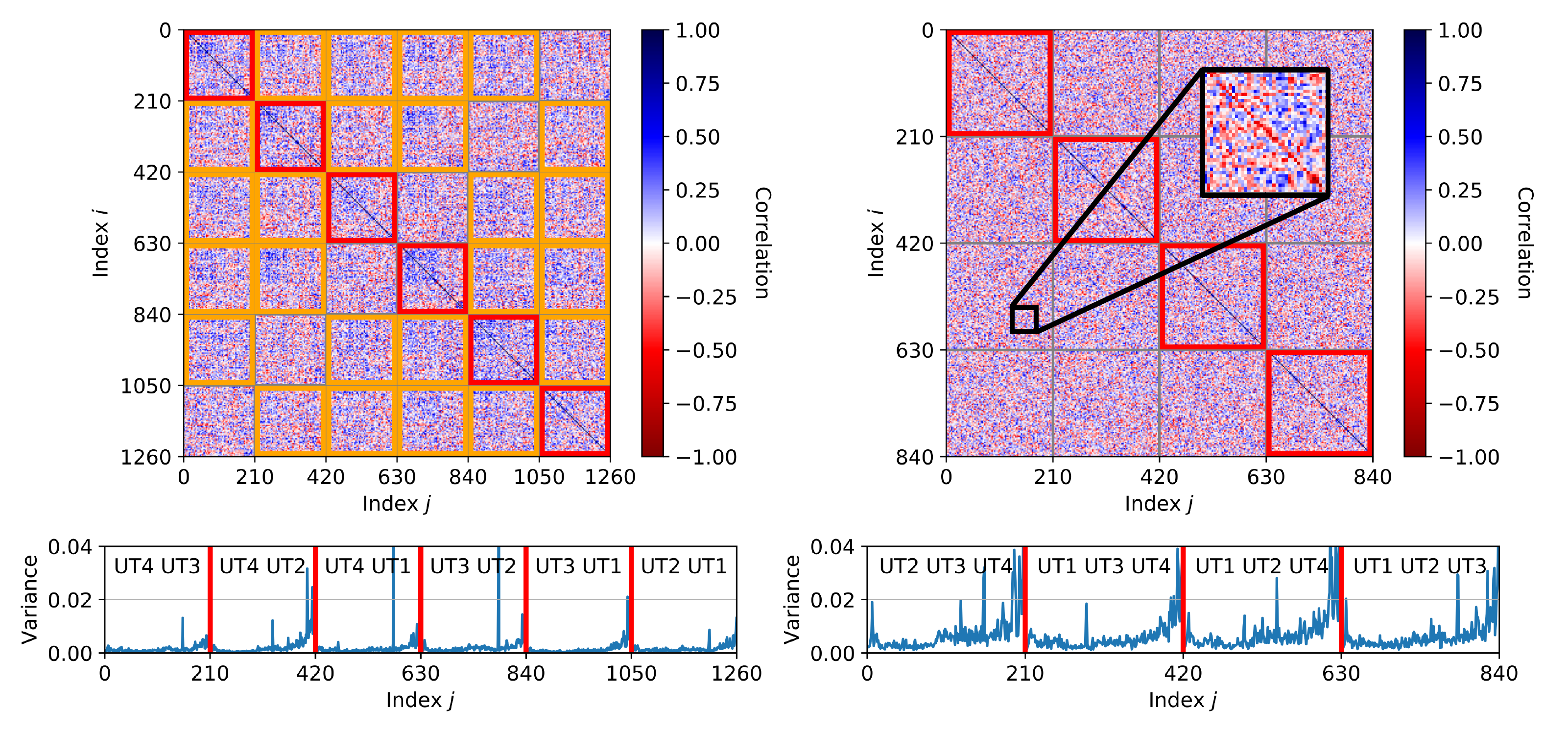}
\includegraphics[width=0.8\textwidth]{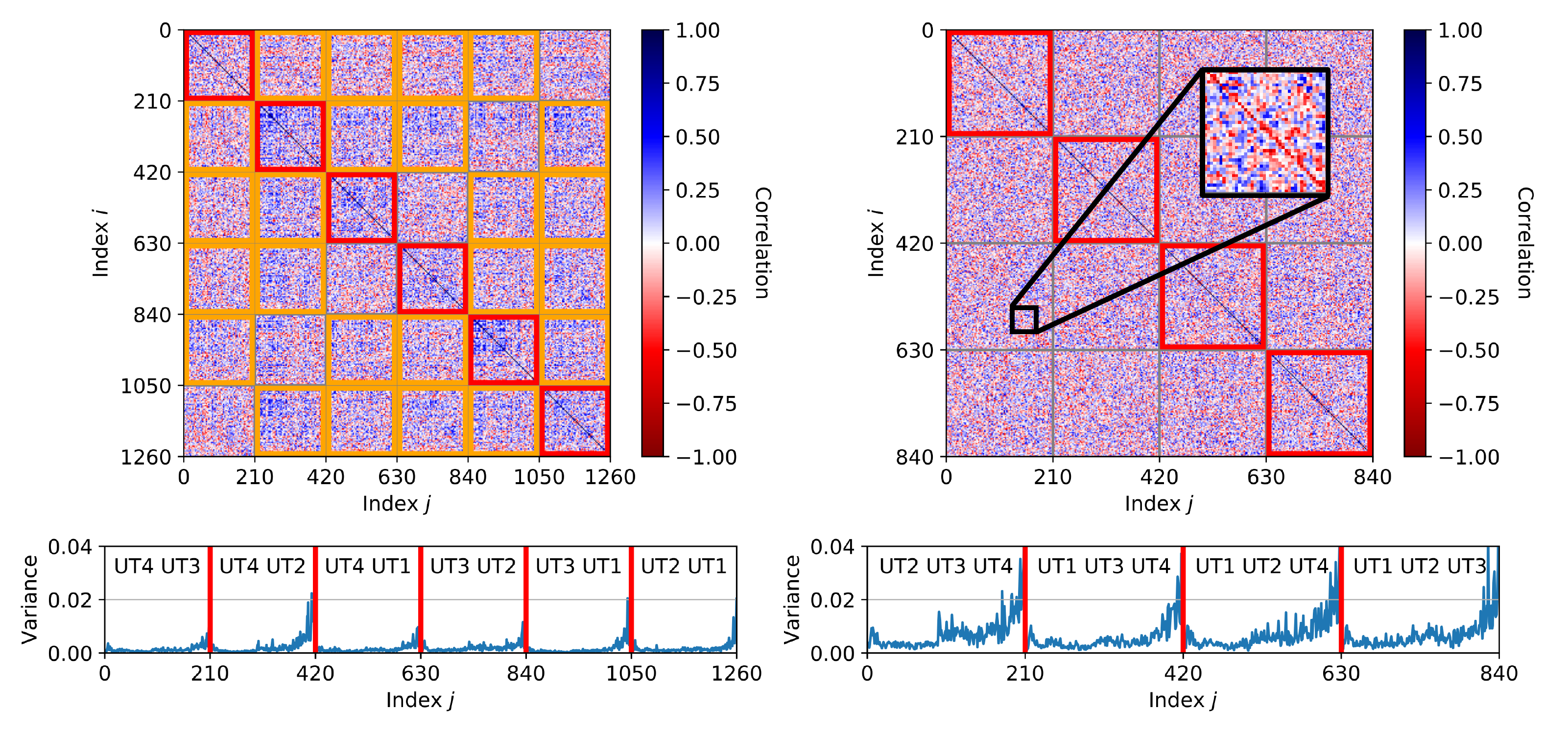}
\end{center}
\caption{Same as Figure~\ref{fig:corr_SC}, but showing the correlations of the VISAMP instead of the VIS2, extracted from the three P2VM-reduced files used for the injection and recovery tests with real data in Section~\ref{sec:injection_and_recovery_tests_real_data} (programme 0101.C-0907(B)). From top to bottom: GRAVI.2018-04-18T08-08-19.739\_singlescip2vmred.fits, GRAVI.2018-04-18T08-12-10.749\_singlescip2vmred.fits, GRAVI.2018-04-18T08-20-04.769\_singlescip2vmred.fits.}
\label{fig:y_corr_SC}
\end{figure*}

\section{Linearised model}
\label{app:linearised_model}

If the host star is essentially unresolved (i.e. $\theta b\lambda \ll 1$) and the companion is at high contrast (i.e. $f \ll 1$) one can linearise the $\text{VIS2}_\text{bin}$ and the $\text{T3}_\text{bin}$ as a function of the relative flux of the companion.

Consider the complex visibility of the binary model $\text{VIS}_\text{bin}$ in the case where $\theta \rightarrow 0 \Leftrightarrow \text{VIS}_\text{ud} \rightarrow 1$ and $f \ll 1$, then the VIS2 of this model is
\begin{align}
    |\text{VIS}_\text{bin}|^2 &\approx \frac{1}{(1+f)^2}\left[\left(1+f\cos(x)\right)^2+f^2\sin^2(x)\right] \\
    &= \frac{1}{(1+f)^2}\left[1+2f\cos(x)+f^2\right]\\
    &= \left(1-2f+\mathcal{O}(f^2)\right)\left[1+2f\cos(x)+f^2\right] \\
    &= 1+2f\cos(x)-2f+f^2-4f^2\cos(x)+\mathcal{O}(f^3) \\
    &= 1+f\left(2\cos(x)-2\right)+\mathcal{O}(f^2)
\end{align}
and the phase (or argument) of this model (and therefore any linear combination of phases such as the T3) is
\begin{align}
    \angle\text{VIS}_\text{bin} &= \arctan\frac{\mathrm{Im}\text{VIS}_\text{bin}}{\mathrm{Re}\text{VIS}_\text{bin}} \\
    &\approx \frac{\mathrm{Im}\text{VIS}_\text{bin}}{\mathrm{Re}\text{VIS}_\text{bin}} \\
    &= \frac{-f\sin(x)}{1+f\cos(x)} \\
    &\approx f\frac{-\sin(x)}{1}
\end{align}
for $x = -2\pi i(\Delta_\text{RA}u/\lambda+\Delta_\text{DEC}v/\lambda)$. Hence, in the high-contrast regime, one has $\text{VIS2}_\text{bin} \propto 1+f$ and $\text{T3}_\text{bin} \propto f$.

\section{Detection map for real data}

\begin{figure}
\begin{center}
\includegraphics[width=\columnwidth]{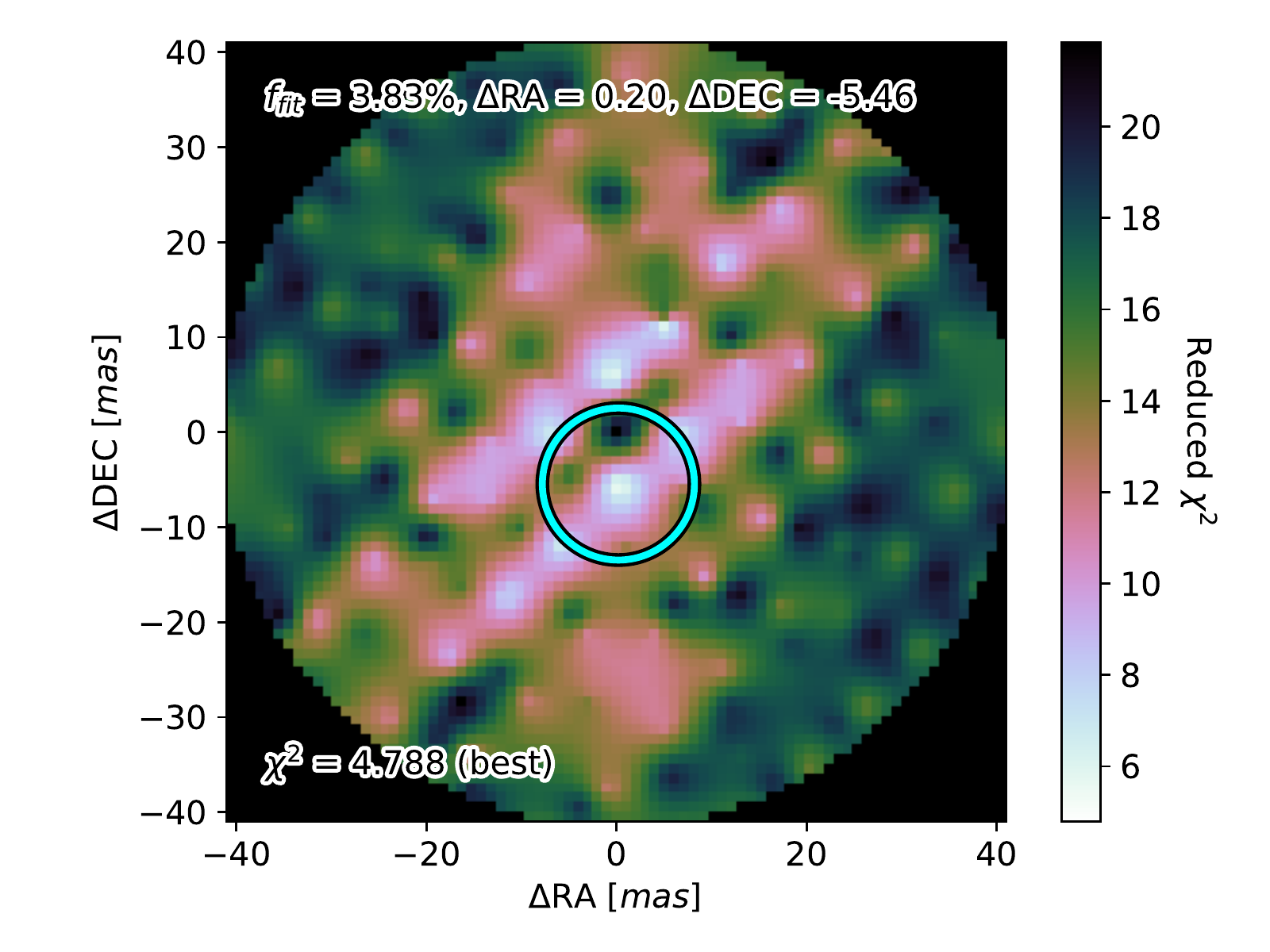}
\end{center}
\caption{Detection map for the GRAVITY data used for the injection and recovery tests in Section~\ref{sec:injection_and_recovery_tests_real_data}. The host star is located in the centre of the map and the cyan circle highlights the position of the best fit companion. Its parameters and reduced chi-squared are shown at the top and the bottom of the map. North is up and east is left.}
\label{fig:detection_map}
\end{figure}

\end{appendix}

\bibliographystyle{aa}
\bibliography{references}

\end{document}